\documentclass[structabstract]{aa}  
\pdfoutput=1
\usepackage{graphicx}
\usepackage{color}
\usepackage{txfonts}
\usepackage{longtable,lscape}
\usepackage{natbib}
\begin{document}
   \title{CH$_{3}$OCH$_{3}$ in Orion-KL: a striking similarity with HCOOCH$_{3}$\thanks{Based on observations carried out with the IRAM Plateau de Bure Interferometer. IRAM is supported by INSU/CNRS (France), MPG (Germany) and IGN (Spain).}}


   \author{N. Brouillet\inst{1,2}     
          \and
          D. Despois\inst{1,2}
          \and 
          A. Baudry\inst{1,2}
          \and 
          T.-C. Peng\inst{1,2}
          \and 
          C. Favre\inst{3}
          \and
          A. Wootten\inst{4}
          \and    
          A. J. Remijan\inst{4}
          \and    
          T. L. Wilson\inst{5}
           \and          
          F. Combes\inst{6}         
          \and
          G. Wlodarczak\inst{7}
          }

  \institute{ Univ. Bordeaux, LAB, UMR 5804, F-33270 Floirac, France\\
              \email{brouillet, despois, baudry@obs.u-bordeaux1.fr}
         \and
            CNRS, LAB, UMR 5804, F-33270 Floirac, France
         \and
         Department of Physics and Astronomy, University of $\AA$rhus, Ny Munkegade 120, DK-8000 $\AA$rhus C, Denmark\\
          \email{favre@phys.au.dk}
          \and
             National Radio Astronomy Observatory, 520 Edgemont Road, Charlottesville, VA 22903-2475, USA\\
             \email{awootten, aremijan@nrao.edu} 
          \and
          Naval Research Laboratory, Code 7210, Washington, DC 20375, USA\\
           \email{tom.wilson@nrl.navy.mil} 
          \and
             Observatoire de Paris, LERMA, CNRS, 61 Av. de l'Observatoire, 75014 Paris, France\\
             \email{francoise.combes@obspm.fr}              
           \and
             Laboratoire de Physique des Lasers, Atomes et Mol\'ecules, Universit\'e de Lille1, UMR 8523, 59655 Villeneuve d'Ascq Cedex, France\\
             \email{georges.wlodarczak@univ-lille1.fr}            
             }

   \date{Received 11 July 2012; accepted  05 December 2012}

 
  \abstract
   {Orion-KL is a remarkable, nearby star-forming region where a recent explosive event has generated shocks that could have released complex molecules from the grain mantles.}
   {A comparison of the distribution of the different complex molecules will help in understanding their formation and constraining the chemical models.}
   {We used several data sets from the Plateau de Bure Interferometer to map the dimethyl ether emission with different arcsec spatial resolutions and different energy levels (from E$\rm_{up}$=18 to 330~K) to compare with our previous methyl formate maps.}
   {Our data show remarkable similarity between the dimethyl ether (CH$_{3}$OCH$_{3}$) and the methyl formate (HCOOCH$_{3}$) distributions even on a small scale (1.8$\arcsec$$\times$0.8$\arcsec$ or $\sim$ 500~AU). This long suspected similarity, seen from both observational and theoretical arguments, is demonstrated with unprecedented confidence, with a correlation coefficient of maps $\sim$ 0.8. }
   { A common precursor is the simplest explanation of our correlation.  Comparisons with previous laboratory work and chemical models suggest the major role of grain surface chemistry and a recent release, probably with little processing, of mantle molecules by shocks. In this case the CH$_{3}$O radical produced from methanol ice would be the common precursor (whereas ethanol, C$_{2}$H$_{5}$OH, is produced from the radical CH$_{2}$OH). The alternative gas phase scheme, where protonated methanol CH$_{3}$OH$_{2}$$^{+}$ is the common precursor to produce methyl formate and dimethyl ether through reactions with HCOOH and CH$_{3}$OH, is also compatible with our data. Our observations cannot yet definitely allow a choice between the different chemical processes, but the tight correlation between the distributions of HCOOCH$_{3}$ and CH$_{3}$OCH$_{3}$ strongly contrasts with the different behavior we observe for the distributions of ethanol and formic acid. This provides a very significant constraint on models.}


   \maketitle
%
%
\section{Introduction}

The Orion protocluster region is the closest high mass star formation region to the Sun. We adopt a distance of  414$\pm$7~pc \citep{Menten:2007}, consistent within the error bars with the values of 389~pc of \citet{Sandstrom:2007}, 437~pc of \citet{Hirota:2007} and 419~pc of \citet{Kim:2008}. This region is remarkable because of the presence of high speed shocks which appear to be generated in the center of the Kleiman-Low infrared nebula Orion-KL. These are reminiscent of a past explosive event and are most clearly seen in the H$_{2}$ 2.12~$\mu$m emission\citep{Allen:1993}. From VLA proper motion measurements of two strong centimetric radiosources, it was proposed that a very unique phenomenon had taken place some 500-1000 years ago: the close encounter, or collision, of two or more rather massive stars. The objects
involved in such a dynamical interaction could have included the Becklin-Neugebauer object (BN) and sources I and n \citep{Gomez:2005,Rodriguez:2005,Goddi:2011,Nissen:2012}. Traces of this explosive pattern have recently been observed in the CO distribution by \citet{Zapata:2009}.

This very recent and energetic event at the heart of the nebula provides unique conditions for the study of a rich interstellar chemistry. Many molecules could have been released from the grain mantles because of dust heating or sputtering by multiple shocks, especially the relatively large molecules  for which no gas--phase--only formation route is satisfactory.

\begin{table*}
\begin{minipage}[t]{17cm}
\caption{Main parameters of the IRAM Plateau de Bure interferometer data sets where transitions of dimethyl ether are detected.}             
\label{Table.dataset_parameters}      
\centering          
\small\addtolength{\tabcolsep}{-1pt} 
\renewcommand{\footnoterule}{}  
\begin{tabular}{c c c c c c c c c }
\hline\hline       
Bandwidth &  HPBW & \multicolumn{2}{c}{Spectral resolution}  & Flux conversion  & \multicolumn{2}{c}{Synthesized beam}  & Pixel size \\
(GHz) &  ($\arcsec$)  & (MHz) & (~km~s$^{-1}$) & (1 Jy~beam$^{-1}$) & ($\arcsec$ $\times$ $\arcsec$) & PA ($\degr$) & ($\arcsec$ $\times$ $\arcsec$) \\
\hline                    
80.502 - 80.574 \footnote{Observations centered on ($\alpha_{J2000}$ = 05$^{h}$35$^{m}$14$\fs$46, $\delta_{J2000}$ = -05$\degr$22$\arcmin$30$\farcs$59) with a velocity of 6 ~km~s$^{-1}$ with respect to the LSR.} &  60 & 0.625 & 2.33 & 4.6~K & 7.63 $\times$ 5.35 & 15 & 0.80 $\times$ 0.80 \\ 
101.178 - 101.717\footnote{Observations centered on  ($\alpha_{J2000}$ = 05$^{h}$35$^{m}$14$\fs$20, $\delta_{J2000}$ = -05$\degr$22$\arcmin$36$\farcs$00) with a velocity of 8 ~km~s$^{-1}$ with respect to the LSR.}  &  50  & 0.625 & 1.85 & 15.8~K & 3.79 $\times$ 1.99 & 22 & 0.50 $\times$ 0.50\\ 
203.411 - 203.483$^{a}$ &  25  & 0.625 & 0.92 & 7.0~K & 2.94 $\times$ 1.44 & 27 & 0.32 $\times$ 0.32 \\
223.408 - 223.941$^{b}$ &  23  & 0.625 & 0.84 & 17.3~K & 1.79 $\times$ 0.79 & 14 & 0.25 $\times$ 0.25 \\ 
\hline                  
\end{tabular}
\end{minipage}
\end{table*}

From a set of twelve Plateau de Bure observations, we have recently analyzed the deuterated methanol isotopologues CH$_{2}$DOH and CH$_{3}$OD  \citep{Peng:2012} and the methyl formate HCOOCH$_3$  \citep{Favre:2011a} emission in Orion KL. The latter work suggests a rather close morphological relation in several places of the Orion KL nebula between excited H$_2$ emission at 2.12~$\mu$m and methyl formate. 

Even though both methyl formate and dimethyl ether are oxygen-bearing molecules of medium complexity and both can be produced from gaseous methanol \citep[e.g.][]{Peeters:2006} and/or from grain surface chemistry \citep[e.g.][]{Garrod:2008},  these two species are spectroscopically different, and their spectra  behave differently with physical temperature. Furthermore, since the early days of dimethyl ether studies \citep{Snyder:1974,Clark:1979} there was a hint of limited departures from LTE for dimethyl ether. Recently, we have carried out  observations at the EVLA of the J$(K_{-1},K_{+1})$ = 6(1,5) - 6(0,6) EE transition of dimethyl ether at 43.4475415~GHz \citep{Favre:2011b} which shows that the distribution is very similar to that of methyl formate.

In this paper we present the analysis of several dimethyl ether lines present in our Plateau de Bure data set. The maps of the CH$_{3}$OCH$_{3}$ emission at different frequencies and energy levels are shown in Sect. \ref{sec:DE}. In Sect. \ref{sec:temperatures} we compare the spectra at the five main emission positions to synthetic spectra based on the temperatures derived from our previous methyl formate analysis. Finally, we discuss in Sect. \ref{sec:discussion} the similarity in spatial structure and velocity between our high resolution maps of dimethyl ether and methyl formate and examine its implication on the different chemical models of formation of these species, in particular in comparison also with our ethanol and formic acid maps.

%
%
%
\section{Observations}
 \label{sec:Observations}

The data set consists in twelve data cubes obtained with the IRAM Plateau de Bure Interferometer, the parameters of which are presented in Favre et al. (2011a; see their Table 1). We used the GILDAS package\footnote{http://www.iram.fr/IRAMFR/GILDAS} for data reduction. The continuum emission was subtracted in the data cubes by selecting line-free channels as judged by careful visual inspection, discarding any contaminated channels. Finally, we cleaned the data cubes, channel by channel, using the B. G. Clark method \citep{Clark:1980}.

Table \ref{Table.dataset_parameters} presents the parameters of the four cubes where transitions of dimethyl ether are detected. Spatial resolution ranges from 1.79$\arcsec$ $\times$ 0.79$\arcsec$ to 7.63$\arcsec$ $\times$ 5.35$\arcsec$ and spectral resolution from 0.84 to 2.33~km~s$^{-1}$.

\begin{table}
\begin{minipage}[t]{8.5cm}
\caption{Detected and blended transitions of dimethyl ether observed with the Plateau de Bure Interferometer toward Orion-KL. }             
\label{Freq}      
\centering        
\renewcommand{\footnoterule}{}  
\begin{tabular}{llccl}
\hline\hline       
Frequency &Transition & E$\rm_{up}$ & S$\mu$$^{2}$ &Note  \footnote{D: detected. B: blended. PB: partial blend.} \\
 (MHz) &  & (K) & (D$^{2}$) & \\
\hline                
80536.3550 & 5( 2, 3)- 5( 1, 4) AE & 19.3 & 32.2 & D \\
80536.4060 & 5( 2, 3)- 5( 1, 4) EA & 19.3 & 21.5 & D\\
80538.6540 & 5( 2, 3)- 5( 1, 4) EE & 19.3 & 86.0 & D\\
80540.9280 & 5( 2, 3)- 5( 1, 4) AA & 19.3 & 53.7 & D\\
80651.6720 & 21( 5,17)-20( 6,14) AA & 245.8 & 28.4 & B \\
101559.3870    & 12( 2,10)-11( 3, 9) AA &  77.6 &  19.6 & D \\
101559.9530   &  22( 5,17)-21( 6,16) AA &  265.9 &  30.6 & D\\
101561.3170   &  22( 5,17)-21( 6,16) EE &  265.9 &  81.5 & D\\
101562.1200    &  12( 2,10)-11( 3, 9) EE &  77.6 &  52.3 & D\\
101562.6290   &  22( 5,17)-21( 6,16) AE &  265.9 &  10.2 & D\\
101562.7350    &  22( 5,17)-21( 6,16) EA &  265.9 &  20.3 & D\\
101564.8330    &  12( 2,10)-11( 3, 9) AE &  77.6 &  6.5 & D\\
101564.8730    &  12( 2,10)-11( 3, 9) EA &  77.6 &  13.1 & D\\  
203364.0480    &  3( 3, 0)- 2( 2, 0) EA &  18.1 &  5.1 & B\\
203374.1550   &  3( 3, 1)- 2( 2, 0) EE &  18.1 &  28.1 & B\\   
203375.7530    &  3( 3, 1)- 2( 2, 0) AE &  18.1 &  8.4 & B\\
203383.0690    &  3( 3, 1)- 2( 2, 0) AA &  18.1 &  25.3 & B\\
203384.3710    &  3( 3, 0)- 2( 2, 0) EE &  18.1 &  39.3 & B\\
203384.4500    &  3( 3, 1)- 2( 2, 0) EA &  18.1 &  11.8 & B\\
203402.7050    &  3( 3, 0)- 2( 2, 1) EA &  18.1 &  11.8 & B\\
203410.1000	&  3( 3, 1)- 2( 2, 1) EE &  18.1 &  39.3 & PB\\
203411.4020    &  3( 3, 0)- 2( 2, 1) AE &  18.1 &  25.3 & PB\\
203418.7180    &  3( 3, 0)- 2( 2, 1) AA &  18.1 &  42.2 & PB\\
203420.3160    &  3( 3, 0)- 2( 2, 1) EE & 18.1  &  28.2 & PB\\
203423.1070    &  3( 3, 1)- 2( 2, 1) EA &  18.1 &  5.1 & PB\\
223409.4660    &  26( 2,24)-26( 1,25) EE &  330.4 &  276.7 & D\\
223412.0350    &  26( 2,24)-26( 1,25) AA &  330.4 &  103.8 & D\\

\hline                  
\end{tabular}
\end{minipage}
\end{table}

%
%
\section{Dimethyl ether (CH$_{3}$OCH$_{3}$) frequencies and maps}
\label{sec:DE} 
   \begin{figure}[h!]
  \centering
\includegraphics[width=8.8cm]{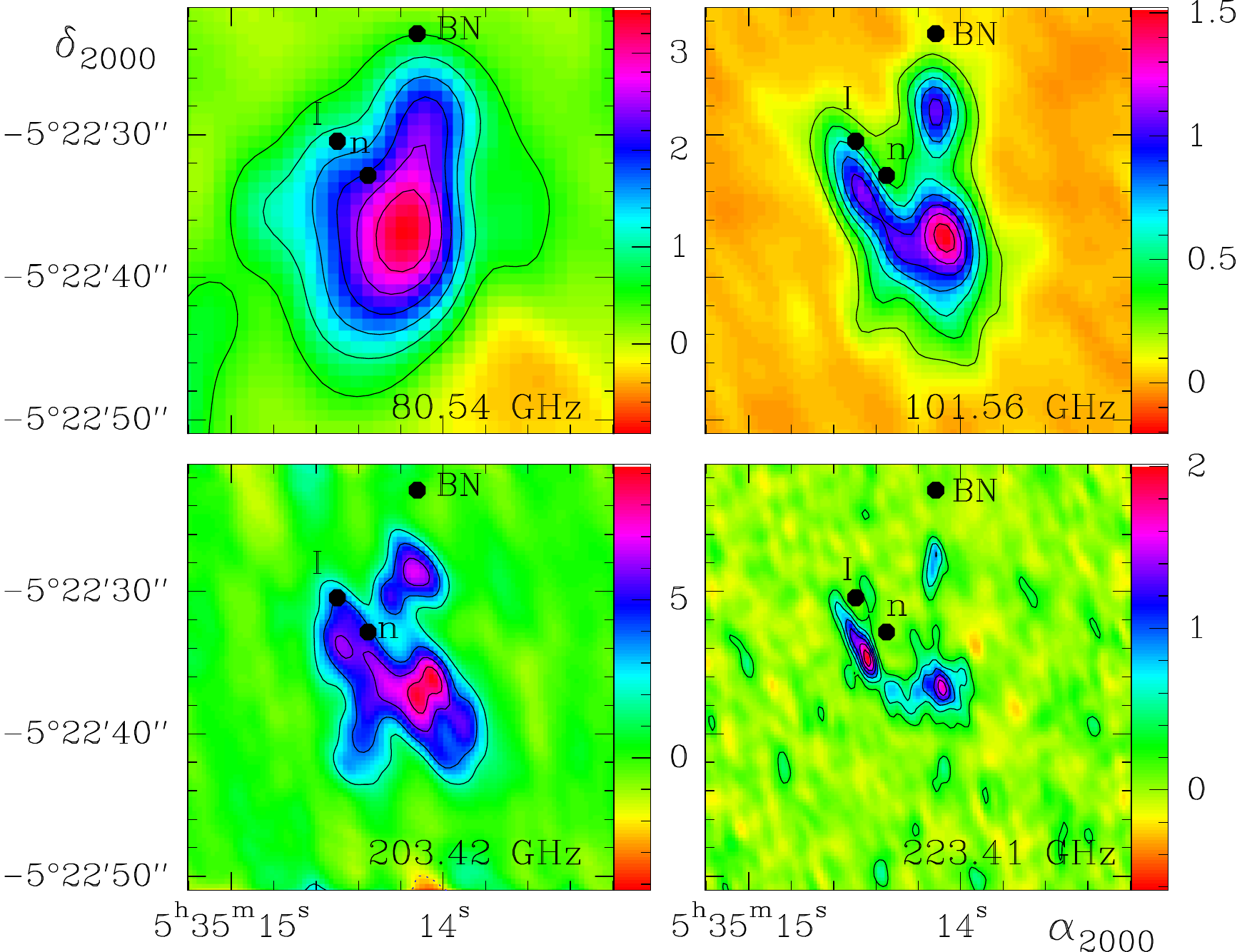}
   \caption{Dimethyl ether integrated intensity maps  obtained with the Plateau de Bure Interferometer: these panels show the sum of  the emission of the transitions at 80.54~GHz, 101.56~GHz, 203.42~GHz and 223.41~GHz. The first  contour and the step are 0.5~Jy/beam, 0.2~Jy/beam, 2~Jy/beam and 0.3~Jy/beam at 80.54~GHz, 101.56~GHz, 203.42~GHz and 223.41~GHz respectively. Note that the 203.42~GHz transitions are blended with the H$_{2}$$^{18}$O line at 203.407~GHz around ($\alpha_{J2000}$ = 05$^{h}$35$^{m}$14$\fs$5, $\delta_{J2000}$ = -05$\degr$22$\arcmin$34$\arcsec$). The BN object position is ($\alpha_{J2000}$ = 05$^{h}$35$^{m}$14$\fs$1094, $\delta_{J2000}$ = -05$\degr$22$\arcmin$22$\farcs$724), the radio source I position is ($\alpha_{J2000}$ = 05$^{h}$35$^{m}$14$\fs$5141, $\delta_{J2000}$ = -05$\degr$22$\arcmin$30$\farcs$575), and the IR source n position is ($\alpha_{J2000}$ = 05$^{h}$35$^{m}$14$\fs$3571, $\delta_{J2000}$ = -05$\degr$22$\arcmin$32$\farcs$719) from  \citet{Goddi:2011}.  }
              \label{fig1-cartes}%
    \end{figure}
   \begin{figure}[h!]
  \centering
   \includegraphics[width=8.5cm]{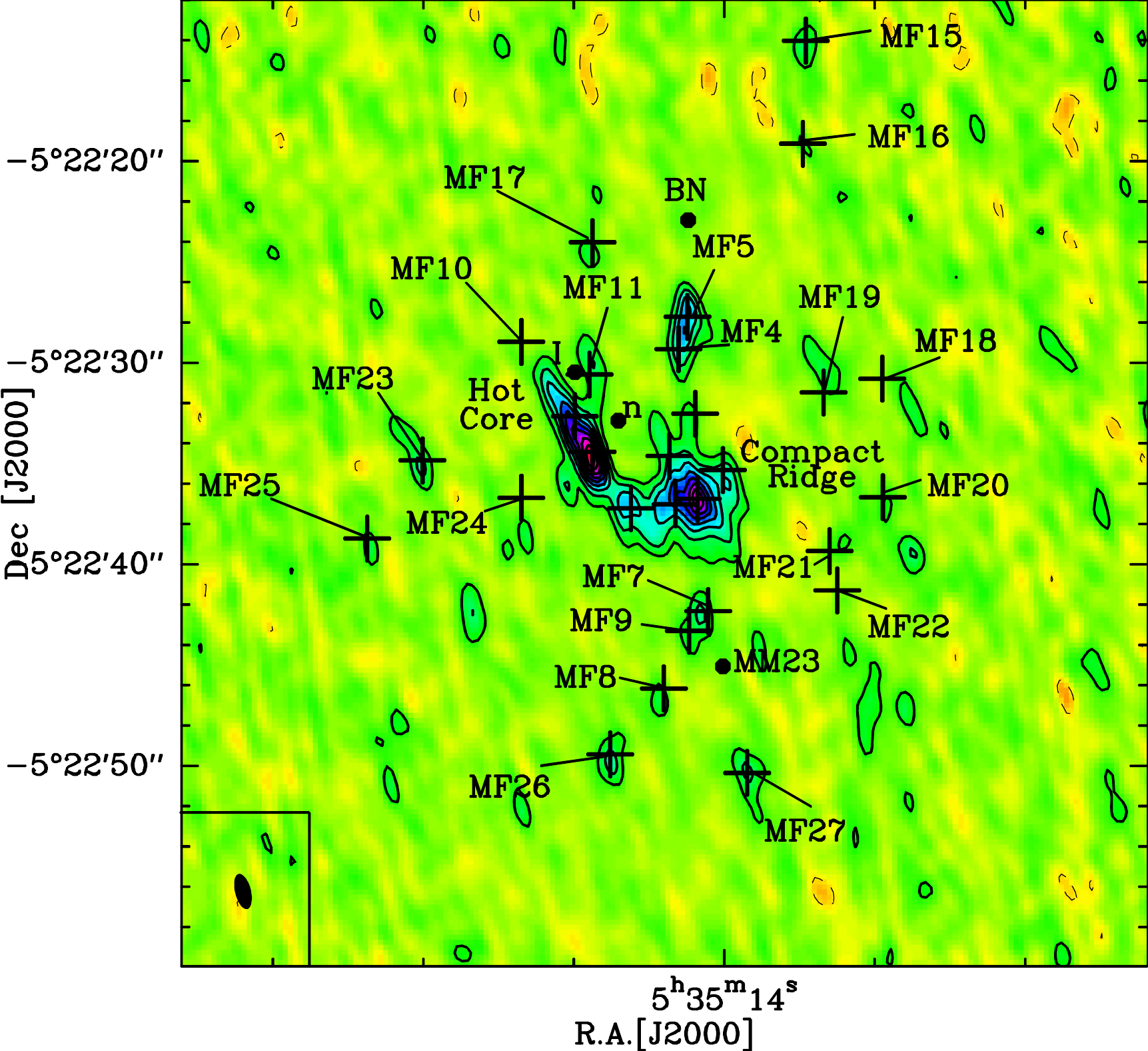}
      \includegraphics[width=8cm]{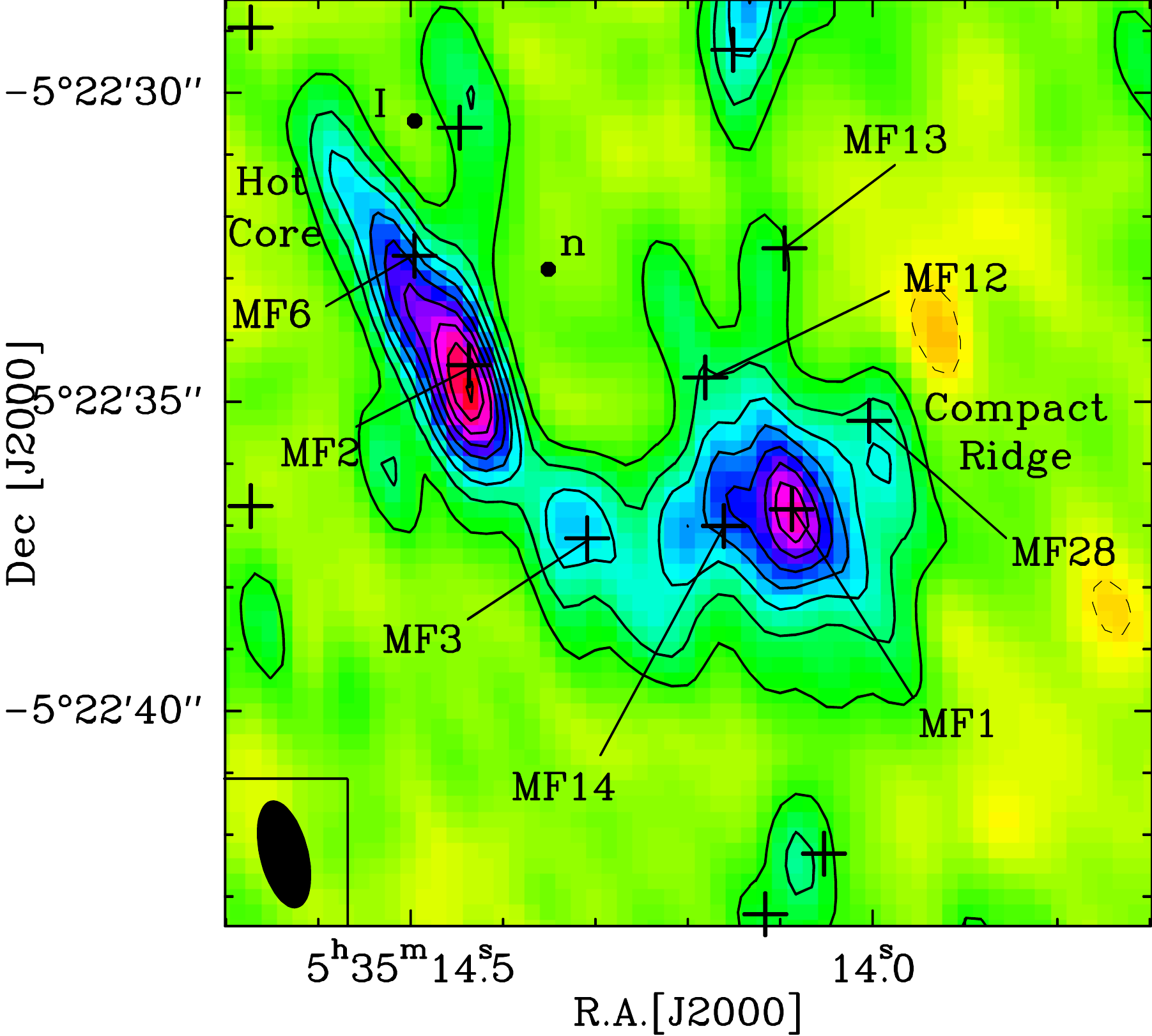}
   \caption{Dimethyl ether integrated intensity maps  obtained with the Plateau de Bure Interferometer (sum of  emission at 223.409~GHz  and 223.412~GHz between 5 and 12~km~s$^{-1}$). The bottom image is a blowup of the Hot Core/Compact Ridge map area. The beam is 1.79$\arcsec$ $\times$ 0.79$\arcsec$; the level step and first contour are 3.8~K~km~s$^{-1}$. The position of the millimeter source MM23  \citep{Eisner:2008} is also indicated. The main different HCOOCH$\rm_{3}$ emission peaks identified in \citet{Favre:2011a} are marked by a cross and labeled MF$\rm_{NUMBER}$. Note that the dimethyl ether distribution is very similar to that of methyl formate \citep[cf. Fig. 4 in][and see Sect. \ref{sec:similarity}]{Favre:2011a}.}
              \label{fig-carte}%
    \end{figure}

\subsection{Dimethyl ether frequencies}

The data set includes several dimethyl ether (DME) lines. Table \ref{Freq} lists the DME transitions, present in our data and detected or blended, taken from the CDMS database \footnote{http://www.astro.uni-koeln.de/cdms} \citep{Muller:2001,Muller:2005} up to E$\rm_{upper}$ $\la$ 650~K and based on the recent work of  \citet{Endres:2009}. The non detected transitions correspond to lines too faint to be detected (S$\mu$$^{2}$$\geq$6.5 for the detected transitions and S$\mu$$^{2}$$\leq$4.1 for non detected). The detected transitions cover the energy range 19 to 330~K.

Dimethyl ether is an asymmetric top molecule with two equivalent methyl groups undergoing large amplitude motions along the CO-bond. The two internal rotations cause a splitting of each rotational level into four substates AA, EE, EA, and AE (see Sect. \ref{sec:temperatures} for line strengths and frequency separation of the multiplets).

%
%
\subsection{The dimethyl ether emission maps}
\label{sec:DMEmaps}

The maps of the dimethyl ether molecule, CH$_{3}$OCH$_{3}$, allow us to trace the spatial distribution of one major oxygenated molecule in Orion-KL. Figure \ref{fig1-cartes} shows the CH$_{3}$OCH$_{3}$ emission measured at different wavelengths; note the different angular resolutions of the various data sets (listed in Table \ref{Table.dataset_parameters}). The distribution shows the same extended, V-shaped molecular emission linking the radio source I to the BN object as previously observed in methyl formate \citep{Favre:2011a}. The highest spatial resolution map at 223.41 GHz (1.79$\arcsec$$\times$0.79$\arcsec$) is presented in Fig. \ref{fig-carte}. We have marked the Hot Core and Compact Ridge positions \citep[see e.g.][]{Beuther:2005}.  The position of the main emission peaks identified in methyl formate, labeled MF1 to MF28, are indicated. Most of these correspond to the dimethyl ether peaks as well within the beam size.

The distribution of CH$_{3}$OCH$_{3}$ appears very similar to the distribution of HCOOCH$\rm_{3}$ in the velocity range where the main component lies between 6 and 9~km~s$^{-1}$ and a north-south linear structure shows up at higher velocities (9--12~km~s$^{-1}$).

\begin{figure}[h!]
  \centering
\includegraphics[width=8.8cm]{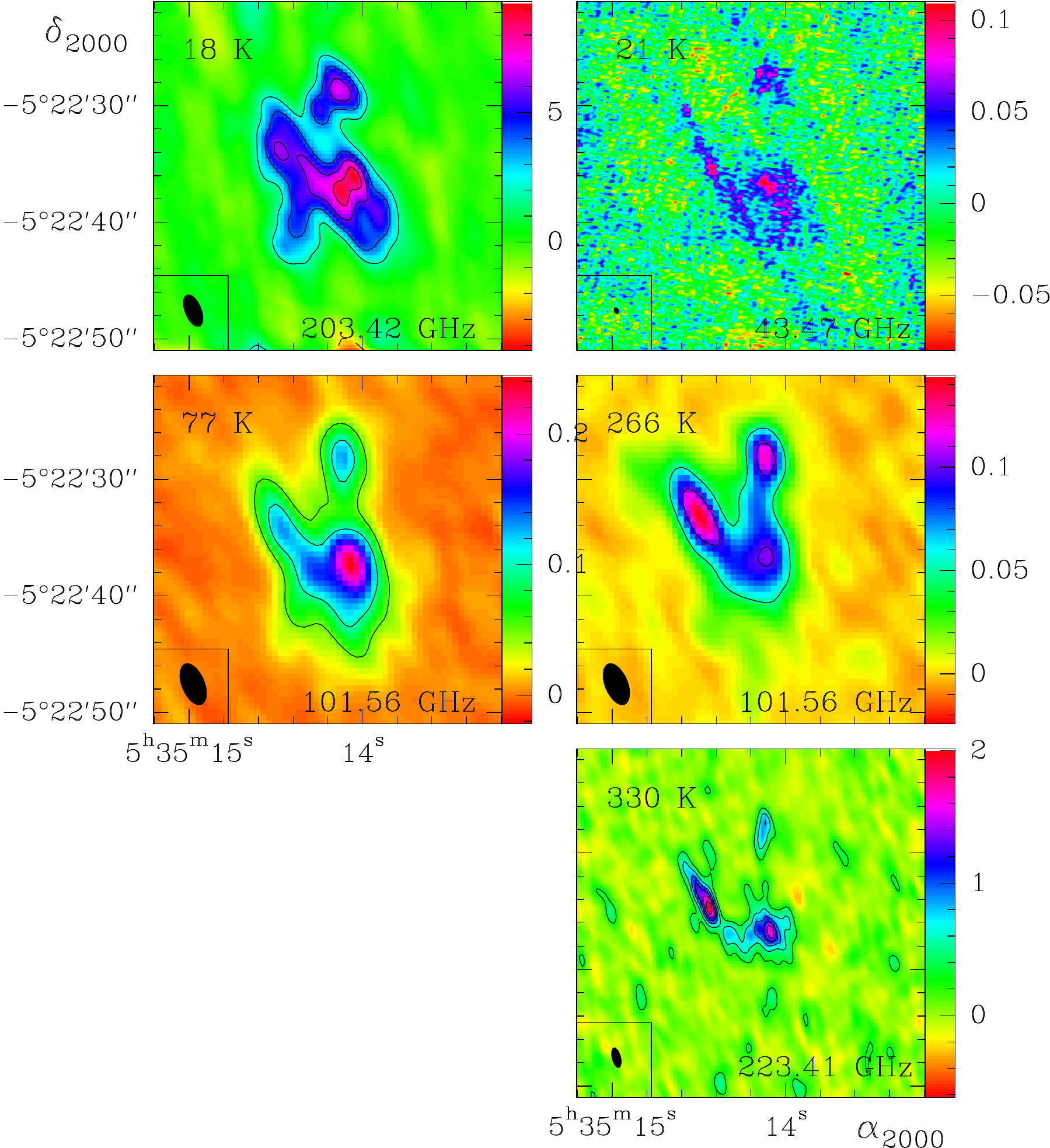}
   \caption{CH$_{3}$OCH$_{3}$ intensity maps integrated in velocity between 5 and 12~km~s$^{-1}$. The line frequency and the upper state energy are indicated on each plot.  The CH$_{3}$OCH$_{3}$ 43.47~GHz map was obtained with the EVLA  \citep{Favre:2011b}. The dimethyl ether emission is stronger towards the compact ridge than the hot core SW position for low upper energy transitions, while the opposite is the case for high upper state energies.}
              \label{fig-temp}%
    \end{figure}

The relative intensities of the different spatial emission peaks depend on the upper state energy of the transitions. Figure \ref{fig-temp} shows the CH$_{3}$OCH$_{3}$ emission for transitions with different energy levels ranging from E$\rm_{up}$=18 to 330~K. The 43.47~GHz map is that obtained with the EVLA \citep{Favre:2011b}. As previously observed for the methyl formate in \citet{Favre:2011a}, the emission of dimethyl ether at the MF2 position \citep["Hot Core SW" in][]{Friedel:2008} becomes stronger at higher energy levels.

\begin{figure}[h!]
  \centering
   \includegraphics[width=5cm,angle=270]{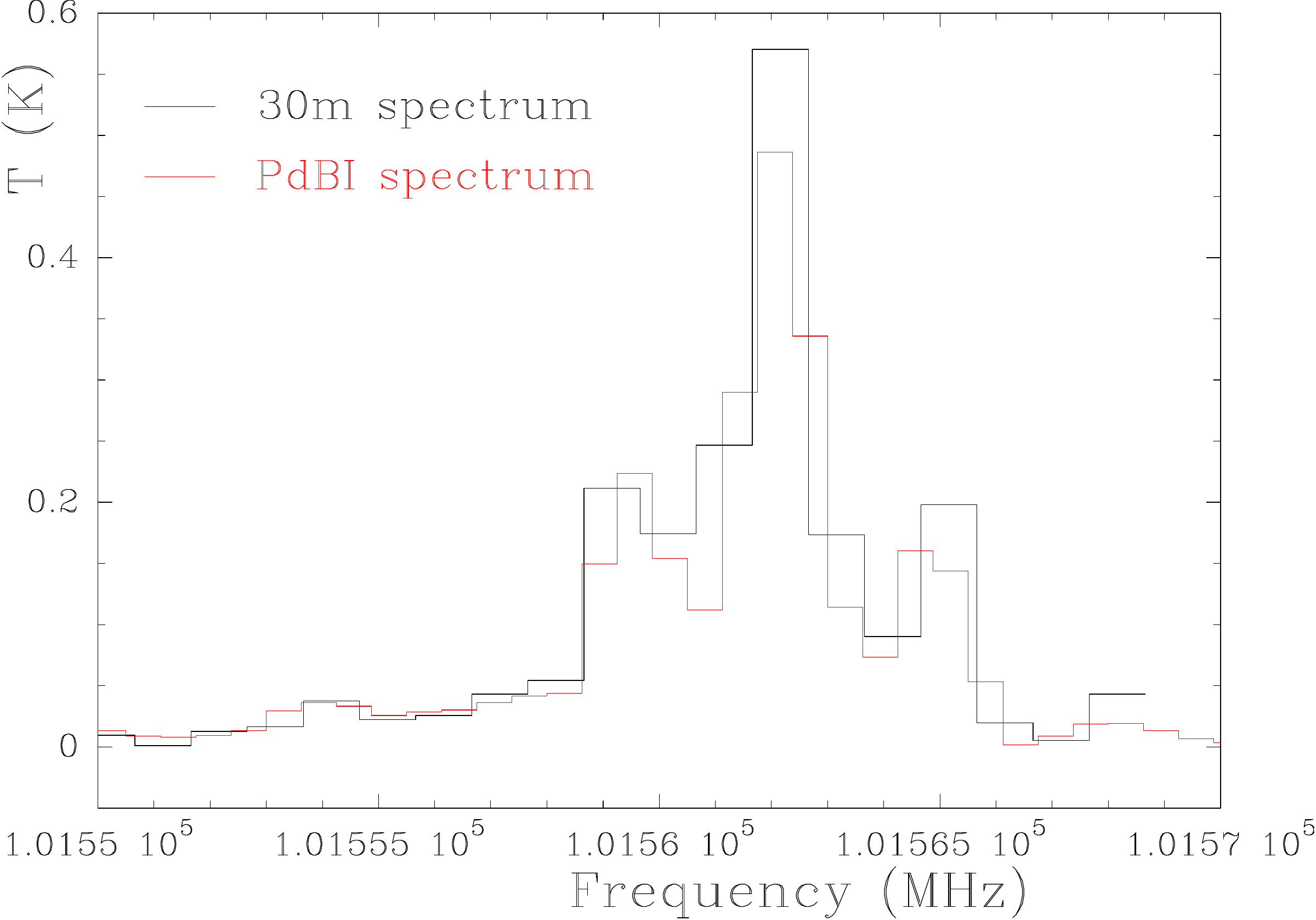}
   \caption{Dimethyl ether spectrum at 101.56~GHz observed with the IRAM 30m radiotelescope (in black, J. Cernicharo, private comm.) overlaid on the PdBI spectrum convolved to the same spatial resolution (in red).}
              \label{fig-comp-30m}%
    \end{figure}

The comparison of our interferometric spectrum at 101.56~GHz to the single-dish spectrum taken with the IRAM 30m (J. Cernicharo, private comm.) shows that little flux is missing due to filtering on a 3$\arcsec$ scale (see Fig. \ref{fig-comp-30m}).


\begin{table}
\caption{Positions of the main HCOOCH$\rm_{3}$ emission peaks observed with the Plateau de Bure Interferometer toward Orion-KL as identified in \citet{Favre:2011a}. These peaks also coincide with the main CH$_{3}$OCH$_{3}$ peaks.}             
\label{Table.position-clumps}      
\centering        
\begin{tabular}{l c c}
\hline\hline       
Position name & R.A (J2000) & Dec (J2000) \\
 & 05$^{h}$35$^{m}$ & -05$\degr$22$\arcmin$ \\
\hline                    
MF1  & 14$\fs$09  & 36$\farcs$7 \\ 
MF2 & 14$\fs$44  & 34$\farcs$4 \\  
MF3  & 14$\fs$31 & 37$\farcs$2 \\ 
MF4  & 14$\fs$15 & 29$\farcs$3 \\ 
MF5  & 14$\fs$12 & 27$\farcs$7 \\ 
\hline                  
\end{tabular}
\end{table}

%
%
\section{Temperatures and column densities}
\label{sec:temperatures}

The four torsional forms AA, EE, AE and EA have weights of 6:16:2:4 for ee, oo levels and 10:16:4:6 for eo and oe levels \citep{Turner:1991,Lovas:1979}. 
Most of the time the spectral resolution of the radiotelescopes does not allow separation of the AE and EA transitions, so that a symmetrical triplet is observed. In that case and if the lines are optically thin, the observed dimethyl ether triplet (AA, EE, AE+EA) should then be either 6:16:6 or 10:16:10 in relative intensity.
For part of the triplet observed at the edge of the bandwidth at 223.4~GHz  we find a ratio of 1:2.7:1 within the errors. This is in agreement with a low optical depth. However for the transitions at 80.53~GHz, the EE line in the middle is much too weak with respect to the other lines. This EE line is stronger in the spectra taken with single-dish radiotelescopes \citep{Johansson:1985,Turner:1989}, so that this could be caused by an observational rather than an excitation effect.\\

\begin{figure}[h!]
  \centering
\includegraphics[width=8.8cm]{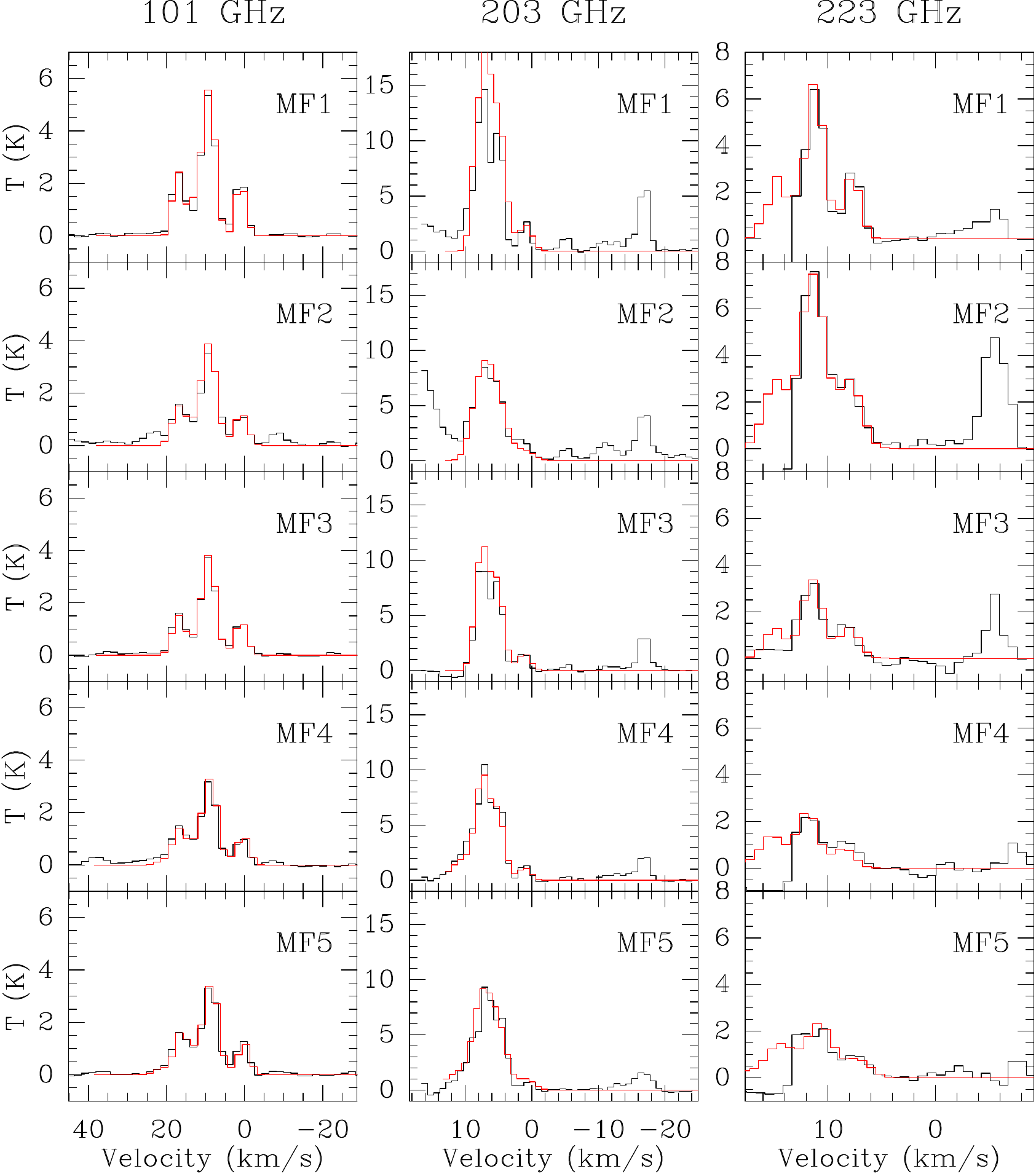}
   \caption{CH$_{3}$OCH$_{3}$ synthetic spectra (in red) overlaid on the observed spectra toward the MF1 to MF5 positions. The 101~GHz,  203~GHz and 223~GHz observations are in the left, middle and right columns respectively. The observed spectrum at 223~GHz is at the edge of the bandwidth.}
              \label{fig-synth}%
    \end{figure}

For the transitions at 101.56, 203.42 and 223.41~GHz, we have produced synthetic spectra assuming local thermodynamic equilibrium (LTE), using the line parameters derived from our methyl formate analysis toward the five main emission peaks whose positions are given in Table \ref{Table.position-clumps}. We first used the HCOOCH$\rm_{3}$ velocity v, line width $\Delta$v$\rm_{1/2}$ and temperature T$\rm_{rot}$ to derive the N$\rm_{CH_3OCH_3}$ column density; then we have slightly adjusted the parameters for a better fit of the dimethyl ether spectra. This was done with our own routines and XCLASS \footnote{This research made use of the myXCLASS program (https://www.astro.uni-koeln.de/projects/schilke/XCLASS), which accesses the CDMS (http://www.cdms.de) and JPL (http://spec.jpl.nasa.gov) molecular data bases.}. Fig. \ref{fig-synth} shows synthetic spectra overlaid on the observed spectra and Table \ref{synth-param} lists the parameters for the angular resolutions of the 101~GHz  (3.63$\arcsec$ $\times$  2.26$\arcsec$) and 223~GHz (1.79$\arcsec$ $\times$ 0.79$\arcsec$) observations. The parameters used for the 203~GHz observations (angular resolution of 2.94$\arcsec$ $\times$  1.44$\arcsec$) are intermediate values. The calculated opacities (estimated from the ratio of the brightness temperature to the rotational temperature) are generally less than 0.2 and at most 0.4 at MF1. 

%

\begin{table*}
\begin{minipage}[t]{16cm}
\caption{CH$_{3}$OCH$_{3}$ synthetic spectra parameters (velocity, line width, beam-averaged temperature and column density)  which reproduce best the PdBI spectra at the emission peaks MF1 to MF5 for the angular resolutions of the 101~GHz  (3.63$\arcsec$ $\times$  2.26$\arcsec$) and 223~GHz (1.79$\arcsec$ $\times$ 0.79$\arcsec$) observations. Two velocities components are clearly distinguished at MF1, MF4 and MF5.}            
\label{synth-param}      
\centering        
\renewcommand{\footnoterule}{}
\begin{tabular}{c c| c | c |c | c |c | c}
\hline\hline       
Emission peaks & v &  \multicolumn{2}{c}{$\Delta$v$\rm_{1/2}$}  & \multicolumn{2}{c}{T$\rm_{rot}$} & \multicolumn{2}{c}{N$\rm_{CH_3OCH_3}$} \\
 & (km~s$^{-1}$)&  \multicolumn{2}{c}{(km~s$^{-1}$)} & \multicolumn{2}{c}{(K)} &  \multicolumn{2}{c}{(10$^{17}~$cm$^{-2}$)}  \\
 &  &  (1.79$\arcsec$ $\times$ 0.79$\arcsec$) & (3.63$\arcsec$ $\times$  2.26$\arcsec$) & (1.79$\arcsec$ $\times$ 0.79$\arcsec$) & (3.63$\arcsec$ $\times$  2.26$\arcsec$) & (1.79$\arcsec$ $\times$ 0.79$\arcsec$) & (3.63$\arcsec$ $\times$  2.26$\arcsec$) \\
 \hline 
MF1& 7.5 & 1.85 & 3.0 & 80 & 120 & 3.45 & 2.7 \\
MF1 & 9.2 & 1.0 & 1.5 & 120 & 170 & 0.15&  0.7 \\  
\hline
MF2 & 7.7  & 2.55 & 4.0 & 130 &  125 & 2.5 & 2.5\\
\hline
MF3 & 7.7 & 2.2 & 3.7 & 90 & 105 & 1.5 & 2.0\\
\hline
MF4 &  8.2--7.6\footnote{8.2~km~s$^{-1}$ at (1.79$\arcsec$ $\times$ 0.79$\arcsec$) and 7.6~km~s$^{-1}$ at  (3.63$\arcsec$ $\times$  2.26$\arcsec$).} & 2.2 & 3.5 &  100 & 100 &  0.8 & 1.5\\ 
MF4 &  11.0 & 3.0 & 4.5 &  100 & 100 &  0.3 & 0.3\\ 
\hline
MF5 & 7.35 & 2.8 & 3.2 & 110 & 110 & 0.85  & 1.65\\ 
MF5 & 11.5 & 3.0 & 5.0 & 110 & 110 & 0.3  & 0.45\\ 
\hline
\hline                  
\end{tabular}
\end{minipage}
\end{table*}

We derived abundances relative to methanol using the column densities of \citet{Peng:2012}. We find values between 0.05 and 0.2. This result assumes that both gas components are coextensive and in LTE at the same temperature, which might be questionable if the difference in the derived temperature between dimethyl ether and methanol is significant. This question is related to the dimethyl ether formation route and is discussed later in Sect. \ref{sec:discussion}.

%
\section{Discussion}
\label{sec:discussion}

\subsection{Similarity of dimethyl ether and methyl formate HCOOCH$_{3}$ maps}

Most maps of molecular emission in Orion KL obtained in previous studies show a huge diversity in morphology. On the contrary we show here the very striking \emph{similarity} which appears between our high resolution maps of dimethyl ether and methyl formate \citep{Favre:2011a}. The high degree of correlation we find  reinforces the need to explain the similarity already noted between these species in previous studies, and probably points towards a similarity of their formation paths, with a common precursor for both species.

First we present the correlation of dimethyl ether with methyl formate observed in this work and previous results for these molecules. We then briefly review the models of gas phase and/or ice mantle chemistry proposed to form these species, concentrating on the most recent works \citep{Laas:2011,Neill:2011,Bisschop:2007,Oberg:2009,Oberg:2010} and examine the compatibility of various hypotheses with the present data, as well as with our ethanol and formic acid maps made from the present data set.

\subsubsection{PdBI maps and correlation diagrams}
\label{sec:similarity}

\begin{figure}[h!]
 \centering
 Ê\includegraphics[width=4.4cm,angle=270]{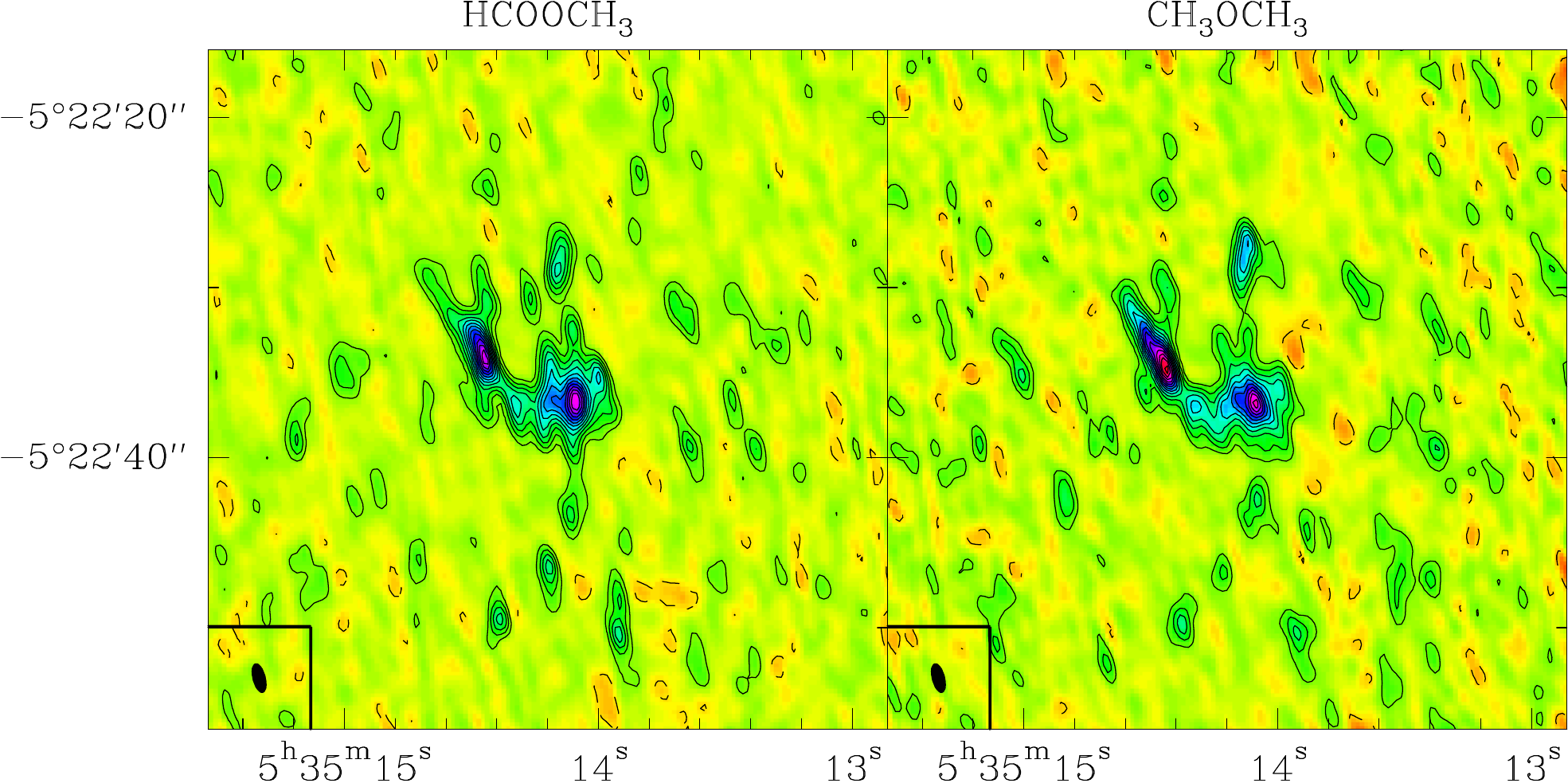}
 Ê\caption{Comparison of the methyl formate map (left) at 223.534~GHz (E$_u$=305~K) and the dimethyl ether map (right) at 223.41~GHz (E$_u$=330~K). The transitions are from the same data cube and the 1.79$\arcsec$ $\times$ 0.79$\arcsec$ beam is plotted in the bottom left corner.}
 ÊÊÊÊÊÊÊÊÊÊÊÊ\label{fig-mf-dme}%
 ÊÊ\end{figure}

When we compare the dimethyl ether map with the methyl formate map obtained with the same data set and from transitions with close E$_u$ energies (Fig. \ref{fig-mf-dme}), these show a striking overall similarity: 1) both molecules are present in the 6-11~km~s$^{-1}$ velocity range and show no detectable emission at 5~km~s$^{-1}$, 2) both molecules show a general V-shaped distribution, 3) they both have their strongest peak at the Compact Ridge for transitions of low E$_u$ energy range (MF1 position), 4) they both have a second peak (MF2 position) at the position named Hot Core SW  by  \citet{Friedel:2008}, a few arcseconds below the continuum Hot Core peak, and which is stronger than MF1 for high E$_u$ energy range transitions (see Sect. \ref{sec:DMEmaps}), 5) they both have a northern extension toward BN around 10~km~s$^{-1}$, 6) many other secondary peaks  are similar. 

The correlation between the two maps is however not 100\%. Faint emission is present for both species but differ somewhat in shape at other places or is displaced by1--2$\arcsec$. One should note however the influence of the cleaning procedure necessary to reconstruct the distribution.

\begin{figure}[h!]
 \centering
 Ê\includegraphics[width=3cm,angle=270]{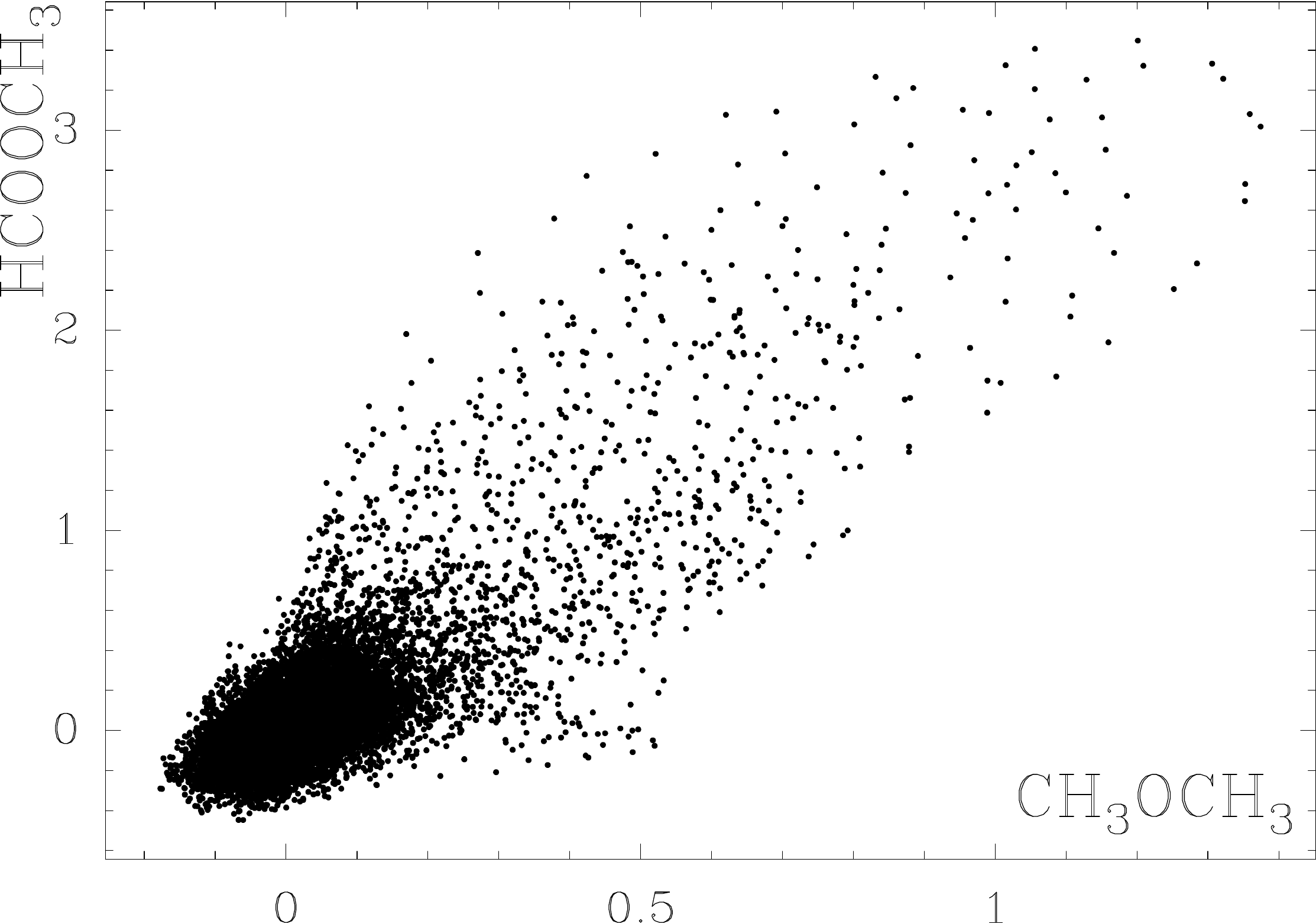}
  Ê\includegraphics[width=3cm,angle=270]{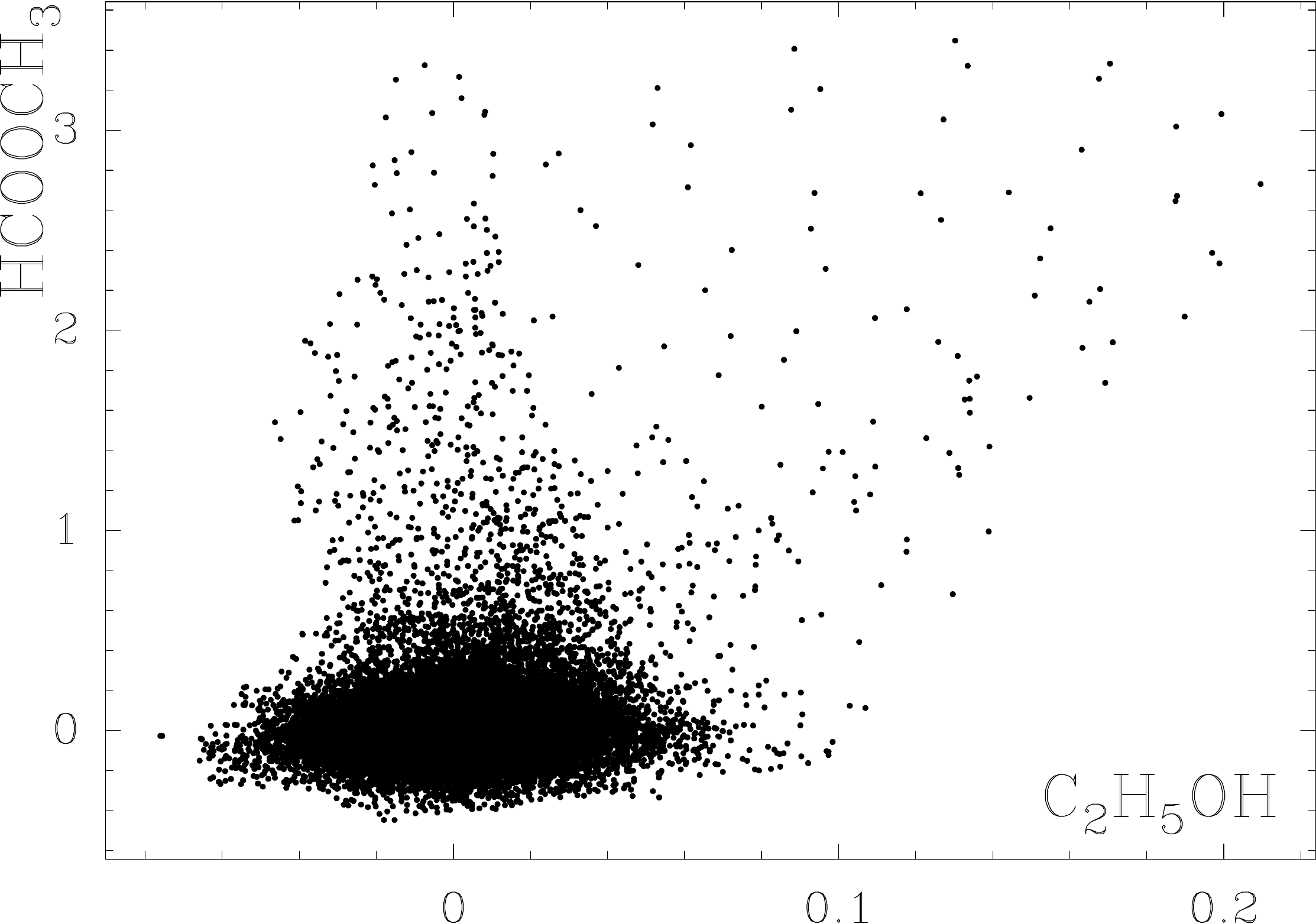}
   Ê\includegraphics[width=3cm,angle=270]{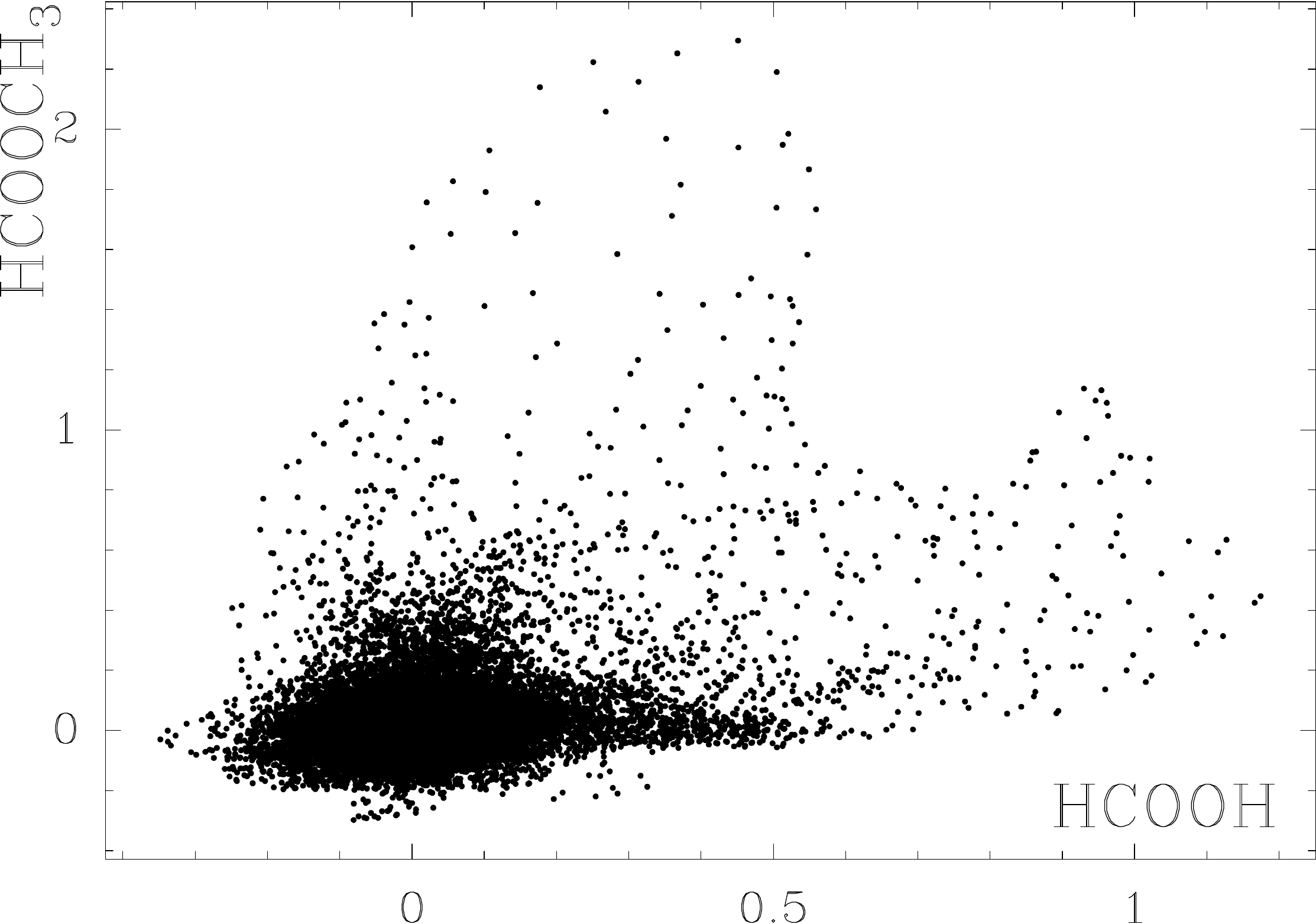}
      Ê\includegraphics[width=3cm,angle=270]{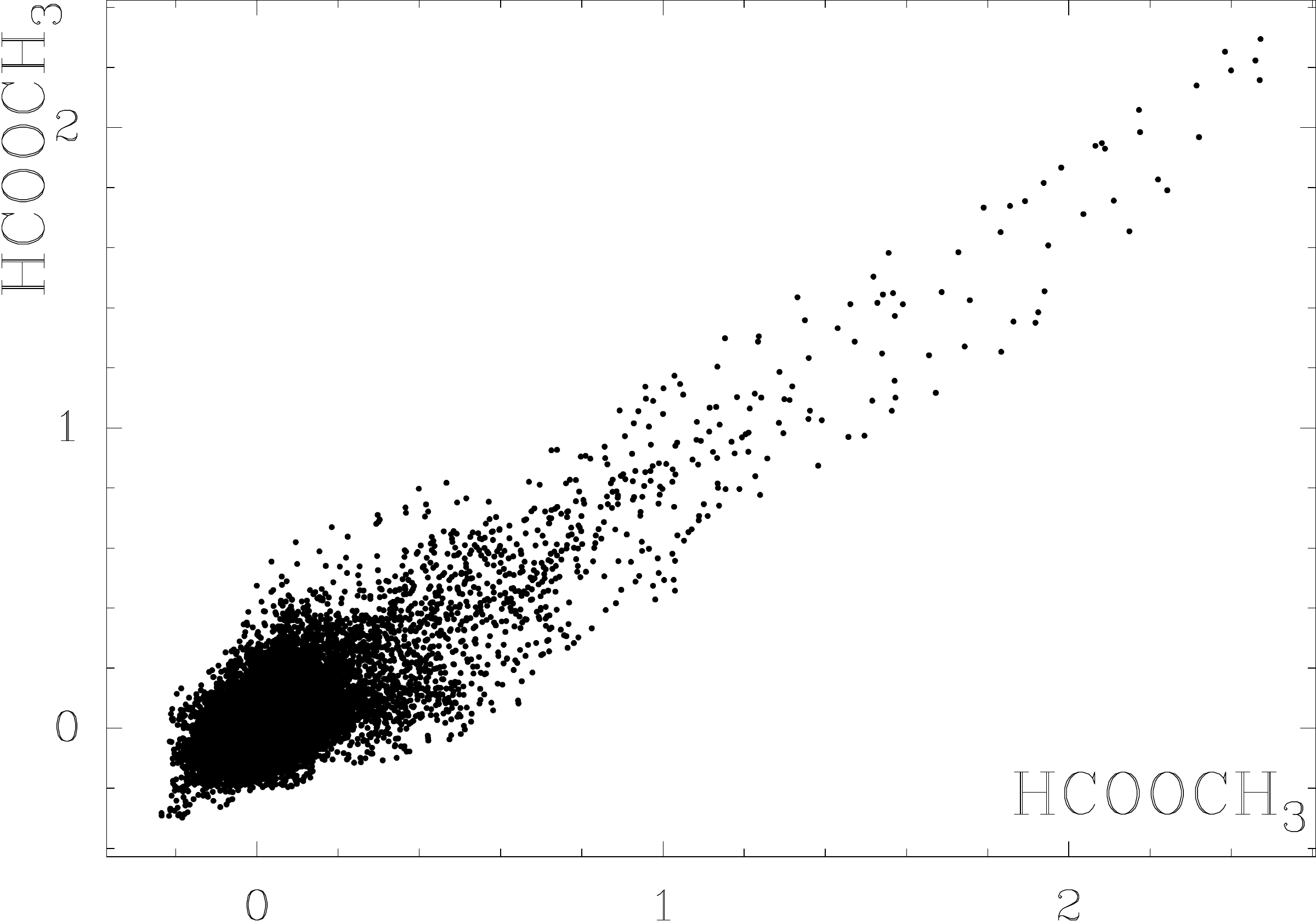}
 Ê\caption{A plot of the intensity of pixels in the methyl formate maps versus the intensity of the same pixels in the dimethyl ether (top left), ethanol (top right), formic acid (bottom left) and methyl formate (bottom right) maps. Care has been taken to use transitions with similar E$_u$ energies. The methyl formate (E$_u$=305~K) at 223.534~GHz is compared to the dimethyl ether transition (E$_u$=330~K) at 223.41~GHz and to the ethanol transition (E$_u$=346~K) at at 223.629~GHz. The methyl formate transition (E$_u$=50~K) at 223.500~GHz is compared to the formic acid transition (E$_u$=72~K) at 223.915~GHz and to the methyl formate transition (E$_u$=50~K) at 223.465~GHz. Dimethyl ether shows clearly the highest spatial correlation with methyl formate. }
 ÊÊÊÊÊÊÊÊÊÊÊÊ\label{correlations}%
 ÊÊ\end{figure}

The similarity between dimethyl ether and methyl formate maps can be displayed in a more quantitative way. In Fig. \ref{correlations} the intensity of each pixel of the dimethyl ether emission line (E$_u$=330~K) at 223.41~GHz and methyl formate emission line (E$_u$=305~K) at 223.534~GHz are compared (note that there is some correlation between pixels since the pixel size is about 3 times smaller than the synthesized beam). The correlation coefficient of Bravais-Pearson is 0.8. By contrast we show the same type of diagram with formic acid HCOOH (10(2,9)-9(2,8) transition at 223.915~GHz with E$_u$=72~K) compared to the methyl formate transition at 223.500~GHz (11(4,8)-10(3,7), E$_u$=50~K) and ethanol C$_{2}$H$_{5}$OH (23(7,17)-23(6,17), v$_t$=0-1 transition at 223.629~GHz with E$_u$=346~K) compared to the methyl formate transition at 223.534~GHz (18(5,14)-17(5,13), E$_u$=305~K), for which the correlation coefficients are only 0.32 and 0.19 respectively. For reference, we also show the correlation of 2 transitions of methyl formate with same energy (223.465~GHz and 223.500~GHz with E$_u$=50~K); the correlation coefficient is 0.82 -- the methyl formate/dimethyl ether result is almost as high. The correlation of 2 transitions of methyl formate with different energies (223.465~GHz and 223.534~GHz with E$_u$=50~K and 305~K,  respectively) is not as good (correlation coefficient of 0.76) due to differences in excitation which are apparent in Fig. \ref{fig-temp} and in \citet[][see their Fig. 15]{Favre:2011a}. The estimated opacity for all the above transitions is less than 0.15, so that there is no noticeable optical depth effect.

This similarity of the dimethyl ether map with the methyl formate map is not observed with other species \citep[see e.g.][and our Fig. \ref{fig-mf-hcooh-2} for formic acid HCOOH]{Guelin:2008}.

%
%
 Ê
%
%
%
 Ê

\subsubsection{Dimethyl ether and methyl formate behavior in previous publications }
\label{sec:previous}

The similarity in the distribution between the two species has also been seen in single dish studies of Orion and of other regions. In Table \ref{previous-studies} we selected only some of the numerous Orion KL line surveys based on two factors: 1) early papers from which the parallel between some species so called "compact ridge species" , and in some cases already the parallel dimethyl ether/methyl formate distribution, was progressively  recognized with increasing evidence, and 2) papers providing the best line profile parameters, temperatures and column densities for both  dimethyl ether and methyl formate. 

\begin{table*}
\begin{minipage}[t]{17cm}
\caption{Dimethyl ether and methyl formate parameters from previous single-dish studies.}             
\label{previous-studies}      
\centering          
\small\addtolength{\tabcolsep}{-1pt} 
\renewcommand{\footnoterule}{}  
\begin{tabular}{l c c c c c c c c c}
\hline\hline       
Authors &  Telescope & Beam & Frequency & \multicolumn{2}{c}{Dimethyl ether}  &  \multicolumn{2}{c}{Methyl formate}  & R \footnote{Normalized number of dimethyl ether molecules, see Sect. \ref{sec:previous}.}  & Comment \\
& &  & &  T$\rm_{rot}$   & N  & T$\rm_{rot}$  & N  &  \\
& & ($\arcsec$) & (GHz) &   (K)  & (10$^{16}$cm$^{-2}$) &  (K) & (10$^{16}$cm$^{-2}$) &  \\
\hline   
\citet{Goddi:2009a}  &  GBT  &  16 &  43 &  88 & 5.8  &  &  & 1.07 & \footnote{Assuming a 10$\arcsec$ source size.} \\                 
\citet{Johansson:1984}  &  Onsala~20m  & 47 & 72-90 & 75 & 0.25 &  25-90 &  0.05-0.25 & 1.02 & \\
\citet{Turner:1991} &  NRAO~11m  &  83 &  75-115 &  91 & 0.13  & 62 &  0.66 & 1.66 & \footnote{a second cooler (T$\rm_{rot}$= 22~K) component is seen in MF only with a lower column density (2.7~10$^{14}$).}\\
\citet{Lee:2002}  & TRAO~14m  & 46 & 138-164 &  130-360 & 0.5-3.6 &  55-120 &  0.6-4.4 & 2-14 & \\
\citet{Remijan:2008} &  NRAO~12m  &  43 &  130-170 &  75 & 0.6  &  &  & 2.05 & \\
\citet{Ziurys:1993}  & FCRAO~14m   &  40 &  150-160 &  109 & 0.3  & 52 & 0.077  & 0.89 & \\
\citet{Blake:1987}  & OVRO~10.4m  & 30 & 215-263 & 63$\pm$5  & 0.3  & 90$\pm$10  & 0.26 & 0.50 & \\
\citet{Sutton:1995}  &  JCMT  & 14 & 334-343 & 75 & 1.49   &  105 &  2.45  &  0.54 &  \footnote{Compact Ridge position.} \\
\citet{Schilke:1997} & CSO   & 20 & 325-360 &  89$\pm$5 & 1.8 &  98$\pm$3 &  1.5 & 1.33 & \\
\citet{Schilke:2001} & CSO   & 11 & 607-725 &  360 & 3 &  316 &  13 & 0.67 & \\
\citet{Comito:2005}  & CSO   & 11 & 795-903 &  160 & 2 &   &   & 0.45 & \\
\citet{Persson:2007}  & Odin satellite  & 2.1$\arcmin$ &  486-492/541-577 & 112 & 13 &  &  & 0.73 & \footnote{Adopted source size 5-6$\arcsec$.} \\
This paper $^{f}$  & PdBI  & 30 &  101/203/223 & 90 & 0.6 &  &  & 1 & \footnote{Our PdBI data convolved to a field of view of 30$\arcsec$.} \\
\hline                  
\end{tabular}
\end{minipage}
\end{table*}

When we convolve our PdBI dimethyl ether data to a typical single-dish field of view of 30$\arcsec$, the spectra can be relatively well fitted with one component of temperature T$\rm_{rot}\sim 90$~K and column density N$\rm_{CH_3OCH_3}\sim 0.6 \times10^{16}cm^{-2}$.

Using M$_X = 4.64 \times 10^{-10}  \times \mu_X \times \frac{N_X}{10^{16} cm^{-2}} \times (\frac{\theta}{\arcsec})^2 \times (\frac{d}{414 pc})^2 $
to convert the column density of molecule X (averaged over a beam of diameter $\theta$) into a  mass M$_X$ (in solar mass M$_\odot$), we find for dimethyl ether ($\mu_X$ = 46): M$_{DME} \sim 8.6 \times 10^{-6}$~M$_\odot$, or about 2.8~M$_{Earth}$. 

To better compare our results with the single dish studies, we have computed the dimethyl ether mass in each case, and we display the value, R, relative to our result, in column 9 of Table \ref{previous-studies}. All values are within a factor of two with respect to ours except in the case of the \citet{Lee:2002} observations. No trend is seen with respect to the beam size, despite the large range of sizes; this implies that at least 50\%, and perhaps almost all of the dimethyl ether seen in the ODIN wide beam (2.1$\arcmin$) is in fact contained in the inner 30$\arcsec$ which we have observed with the PdBI. There is no noticeable effect either related to the range of upper level E$_u$  used in the various studies.

The parallel between both molecules is especially striking on small scales in the present PdBI data. Previous interferometric papers showed some hints of this tendency. \citet{Minh:1993}  present the first HCOOCH$_{3}$ and CH$_{3}$OCH$_{3}$ maps, made with the NMA (5$\farcs$2 $\times$ 4$\farcs$2 resolution); they noted that the difference between the emission distributions of the two molecules may be due to the 100~K difference in upper energy of the observed transition levels. \citet{Blake:1996} with OVRO at 1.3~mm (1$\farcs$5 $\times$ 1$\farcs$0 synthesized beam, and 1~MHz = 1.3~km/s  spectral resolution) observed 18 species including methanol, methyl formate and dimethyl ether. No maps are shown for dimethyl ether, but there is a general trend to behave somewhat like HCOOCH$_{3}$, CH$_{3}$OH, H$_{2}$CS and H$_{2}$CO. \citet{Beuther:2005,Beuther:2006} show many maps obtained with the SMA at 0.86~mm and 0.44~mm of O-bearing species including dimethyl ether, methanol and methyl formate, but at different excitation levels and there is only a partial similarity of the maps. In their CARMA observations at 1.3~mm (2$\farcs$5 $\times$ 0$\farcs$85), \citet{Friedel:2008} observed both dimethyl ether (transition at E$_u$=81~K) and methyl formate (transition at E$_u$=120~K). The maps of the two species did not match perfectly, one peaking at IRc5 and the other at IRc6, but the CH$_{3}$OCH$_{3}$ line is blended and a blend with an N-bearing molecule would explain the observed strong emission at IRc6 and the Hot Core SW. Note also that no methyl formate is attributed to the Compact Ridge \citep[while it is the strongest methyl formate peak in][]{Favre:2011a}; but these authors define the Compact Ridge position differently. The correlation is very clear in Fig. 3 of \citet{Neill:2011} from their CARMA observations; the 6$\farcs$1 $\times$ 5$\farcs$0  resolution  is however insufficient to clearly separate the emission from the many individual methyl formate peaks. 

\subsubsection{The abundance correlation of dimethyl ether with methyl formate}

Another important input is the value of the relative dimethyl ether/methyl formate abundances. Table \ref{abundances} lists the abundances derived at the main positions in our maps for dimethyl ether with respect to methyl formate. Dimethyl ether is about 3--4 times more abundant than methyl formate towards MF3, MF4 and MF5.

Several hypotheses can be considered to explain this correlation:
\begin{itemize}
\item Dimethyl ether and methyl formate would emit at the same location because of a similar excitation of the transitions, i.e. a similar behavior with respect to temperature, density or IR radiation field. But there is no special similarity in the level structure of dimethyl ether and methyl formate, especially as compared to other complex organics like ethanol. 
Furthermore, the physical parameters are not uniform over Orion KL:
the temperature varies from 80 to 160~K for methyl formate, which affects the dimethyl ether/methyl formate line ratio. 
The critical densities are $\sim10^{5-8}$~cm$^{-3}$and $\sim10^{4-7}$~cm$^{-3}$ respectively, while n$\rm_{H_{2}}$ varies from 10$^{4}$ to 10$^{8}$~cm$^{-3}$.
And the IR absorption bands are different, only the CH$_{3}$-O bond is common. 

\item These molecules are just undergoing sublimation at the same temperature. This is likely, but similar abundances in the ices are also required in that case.

\item The similar column densities reflect only the general structure of Orion KL. However dust and other molecules have different patterns.
\end{itemize}

This leaves a correlation of the molecular abundances in the gas phase as the most simple and likely explanation (see further discussion in Sect. \ref{sec:formationDME}).

\begin{table}
\begin{minipage}[t]{8.8cm}
\caption{CH$_{3}$OCH$_{3}$  abundance relative to HCOOCH$_{3}$ at the emission peaks MF1 to MF5 for the angular resolutions of the 101~GHz  (3.63$\arcsec$ $\times$  2.26$\arcsec$) and 223~GHz (1.79$\arcsec$ $\times$ 0.79$\arcsec$) observations. We used the N$\rm_{HCOOCH_3}$ values listed in  \citet{Favre:2011a}.}            
\label{abundances}      
\centering        
\renewcommand{\footnoterule}{}
\begin{tabular}{c  |c | c}
\hline\hline       
Emission peaks &  \multicolumn{2}{c}{N$\rm_{CH_3OCH_3}$/N$\rm_{HCOOCH_3}$} \\
 &   (1.79$\arcsec$ $\times$ 0.79$\arcsec$) & (3.63$\arcsec$ $\times$  2.26$\arcsec$) \\
 \hline 
MF1 \footnote{Component at 7.5~km~s$^{-1}$.} &  2.15 & 9 \\
MF1  \footnote{Component at 9.2~km~s$^{-1}$.} &  0.22 &  4 \\  
\hline
MF2 &  1.6 & 3.5\\
\hline
MF3 &  3.3 & 3.6\\
\hline
MF4 &   & 4\\ 
\hline
MF5 &  & 3.6\\ 
\hline
\hline                  
\end{tabular}
\end{minipage}
\end{table}

\subsubsection{Dimethyl ether and methyl formate observations in other regions}

Single dish observations of several other sources have also noted the parallel between these two species.

In their study of 7 massive hot core regions spread across the Galaxy, \citet{Bisschop:2007} study the statistical correlation between the abundances of 13 organic molecules looking for evidence of grain surface chemistry. Methyl formate and dimethyl ether are mostly seen  where T$\rm_{dust}$ is $>$ 100 K. A strong correlation is found between the abundances of H$_{2}$CO, CH$_{3}$OH, C$_{2}$H$_{5}$OH, HCOOCH$_{3}$ and CH$_{3}$OCH$_{3}$ and the correlation is even $ >$ 0.9 among the last three.  

Comparing the results of \citet{Bisschop:2007} with our findings, we note that 1) if dimethyl ether appears to be strongly correlated with methyl formate, the correlation they observed is similarly high with ethanol, 2) T$\rm_{rot}$ has a large scatter among sources of dimethyl ether, and in some sources may be lower or higher than for methyl formate. Beyond possible observational uncertainties (noise and confusion), one should note the different scales of both studies: while we sample down to (1.79$\arcsec$ $\times$ 0.79$\arcsec$ or $\sim$500~AU) in the nearby Orion region with the PdB interferometer, the single dish results of \citet{Bisschop:2007} are average values for the whole region at Tdust $>$ 100~K which is 1500-5300~AU in radius for their sources - the latter would correspond to a study of Orion over $\sim$ 20$\arcsec$ in diameter, i.e most of the  KL region. This would explain why ethanol is not distinguished from dimethyl ether and methyl formate, and the less tight correlation between dimethyl ether and methyl formate could be due to additional formation pathways/excitation conditions for a fraction of these species on a larger scale. In other words, the process of formation of methyl formate and dimethyl ether we observe in our case, probably linked to shocks, might not appear as ``pure'' and unique on a larger scale.

It is interesting to note that in a recent multi-species interferometric study of the young binary protostar IRAS16293, \citet{Jorgensen:2011} have mapped CH$_{3}$OCH$_{3}$ and HCOOCH$_{3}$. The two molecules appear to be similar, showing two small concentrations of similar intensity around the A and B source components. 

An extensive study of O-bearing species was performed in the Galactic Center clouds by \citet{Requena-Torres:2006,Requena-Torres:2008} which led them to  conclude to a "universal" mantle composition in the various GC clouds; this composition is different in hot corinos \citep{Requena-Torres:2006} and in Hot Cores \citep{Requena-Torres:2008}. They relate the gas phase abundances of these complex organic molecules to sputtering or evaporation due to shocks, a situation similar to some extent to Orion KL. In the 2006 study, 7 lines of methyl formate and 2 of dimethyl ether were observed, with E$_u$/k in the same  20-40~K range; dimethyl ether was not observed in their 2008 study. Note that the rotational temperatures derived in the 2008 study are rather cold ($< $15~K). Due to the large distance of the GC clouds the sources they observe are likely more extended than what we observe in Orion. These authors do not put any emphasis on the dimethyl ether/methyl formate comparison, although the two molecules are noted to have similar abundances with respect to methanol. The grain surface scheme proposed \citep[Fig. 9 in][]{Requena-Torres:2008} does not suggest any close relation between methyl formate and dimethyl ether.

\subsection{Formation of dimethyl ether and methyl formate}
\label{sec:formationDME}

\subsubsection{Chemical models}
Since detailed presentations can be found elsewhere \citep[e.g.][]{Peeters:2006,Neill:2011,Herbst:2009} we summarize here only briefly different models.

A first model of pure gas-phase ion-molecule chemistry was proposed \citep[e.g.][]{Blake:1988} where the protonated methanol ion, CH$_{3}$OH$_{2}$$^{+}$,  produced both CH$_{3}$OCH$_{3}$ and HCOOCH$_{3}$, through reactions with CH$_{3}$OH and H$_{2}$CO respectively, which naturally explained the suspected intimate chemical link between these species.
However, a first difficulty for an interstellar chemistry model is to produce a sufficient amount of these complex species. The pure ion-molecule chemistry proved unable to produce the observed abundances. 
This led to suggest a role of surface chemistry  and of the release of mantle molecules into the gas phase by some process. 
The models considered first injection of water, but injection of methanol itself, processed once in the gas phase by ion-molecule chemistry appeared to be  required. A key species in this scheme remained protonated methanol CH$_{3}$OH$_{2}$$^{+}$ , and a close correlation of methyl formate and dimethyl ether was predicted  \citep[e.g.][]{Charnley:1995, Caselli:1993}. Most of the more complex species predicted to be abundant by the \citet{Charnley:1995} model remain undetected, although one of these (not expected to be the most abundant one), ethyl formate HCOOC$_{2}$H$_{5}$, has recently been detected in Sgr B2 \citep{Belloche:2009}.
This ice mantle methanol release/gas-phase post-processing model could not produce enough methyl formate when calculations by \citet{Horn:2004} showed that the reaction CH$_{3}$OH$_{2}$$^{+}$ + H$_{2}$CO was not possible at interstellar temperatures.

Presently there are two different proposals: 1) new gas-phase ion-molecule formation routes where formaldehyde is replaced by formic acid \citep[e.g.][]{Neill:2011}; 2) a direct formation of methyl formate and dimethyl ether on the grains \citep[e.g.][]{Garrod:2008,Laas:2011}.

\subsubsection{How do these models confront our Orion dimethyl ether and methyl formate data }

In our view, the simplest and most natural explanation of the dimethyl ether-methyl formate correlation is the production of both species by a common precursor reacting with an abundant simple species such as either CH$_{3}$OH, H$_{2}$CO or HCOOH. However, we suggest another - less likely in our view - possibility: the convergence of the dimethyl ether and methyl formate abundances resulting from the long time evolution of many chemical reactions.\\

\paragraph{The common precursor hypothesis: CHO radical (solid phase) vs  CH$_{3}$OH and CH$_{3}$OH$_{2}$$^{+}$ (gas phase)\\}

If solid phase chemistry dominates, the common precursor would be the CH$_{3}$O radical, as detailed in  \"{O}berg et al (2010). In the scheme of UV irradiated ices, photodissociation of CH$_{3}$OH leads to two radicals CH$_{3}$O and CH$_{2}$OH \footnote{The mantle chemistry scheme needs UV radiation to produce radicals. Is it possible in Orion KL? According to our previous study on deuterated methanol \citep{Peng:2012}, the composition of ice mantles might have been largely determined in an earlier phase of the cloud, well before the explosive event. During this earlier phase, the matter was conceivably more uniform and less dense, hence less opaque to UV. The present very high opacity of the cloud \citep[up to A$_{v} >$~1000,][]{Favre:2011a} suggests that processing by external UV sources is now negligible except on a very thin external layer. Inside a dense cloud UV can also be produced by dissociative shocks \citep[e.g.][]{Neufeld:1989} or by secondary electrons \citep[e.g.][]{Gredel:1989,Prasad:1983}.
The presence of at least one B star inside Orion KL  (the BN object) could also be an internal UV source but which is unlikely to reach the compact ridge due to absorption. However, as we have no knowledge of the precise 3D structure of the cloud, we cannot completely exclude that there are holes in the gas and we ignore the 3D dust distribution within Orion-KL. But the simplest origin for the UV mantle irradiation remains that it occurred at an earlier epoch.
Note that radicals can be formed also in mantles also by  irradiation by charged particles \citep[see e.g.][]{Bennett:2007, Herbst:2009}.}. The former is the precursor of methyl formate and dimethyl ether which are formed respectively by reaction with HCO and CH$_{3}$ on the grain. The other radical produced from methanol, CH$_{2}$OH, leads to ethanol C$_{2}$H$_{5}$OH, ethylene glycol CH$_{2}$OHCH$_{2}$OH and glycolaldehyde CH$_{2}$OHCHO. A strong prediction of this model would thus be that the 3 latter molecules should show a somewhat similar distribution, which is likely to be different from that of methyl formate and dimethyl ether. Different conditions, either in the pre-explosion phase (a preencounter role of BN, distance to the Trapezium, different heating, e.g. due to the proximity of source I), or  after the explosive event (direct effect of the shock, or subsequent exposure to the photons of BN, even higher proximity of IR luminous source I)  might explain the different importance of the CH$_{2}$OH and CH$_{3}$O molecules in the Hot Core and in the Compact Ridge. Indeed, our interferometric observations show a similar distribution for C$_{2}$H$_{5}$OH and CH$_{2}$OHCH$_{2}$OH but different from that of HCOOCH$_{3}$ and CH$_{3}$OCH$_{3}$ \citep[][Brouillet et al. in prep.]{Guelin:2008}; however, CH$_{2}$OHCHO is not detected  \citep{Favre:2011a}. Note that some conversion mechanisms between CH$_{2}$OH and CH$_{3}$O are suggested by \citet{Cernicharo:2012} in their CH$_{3}$O detection paper.

\begin{figure}[h!]
 \centering
 Ê\includegraphics[width=7cm,angle=270]{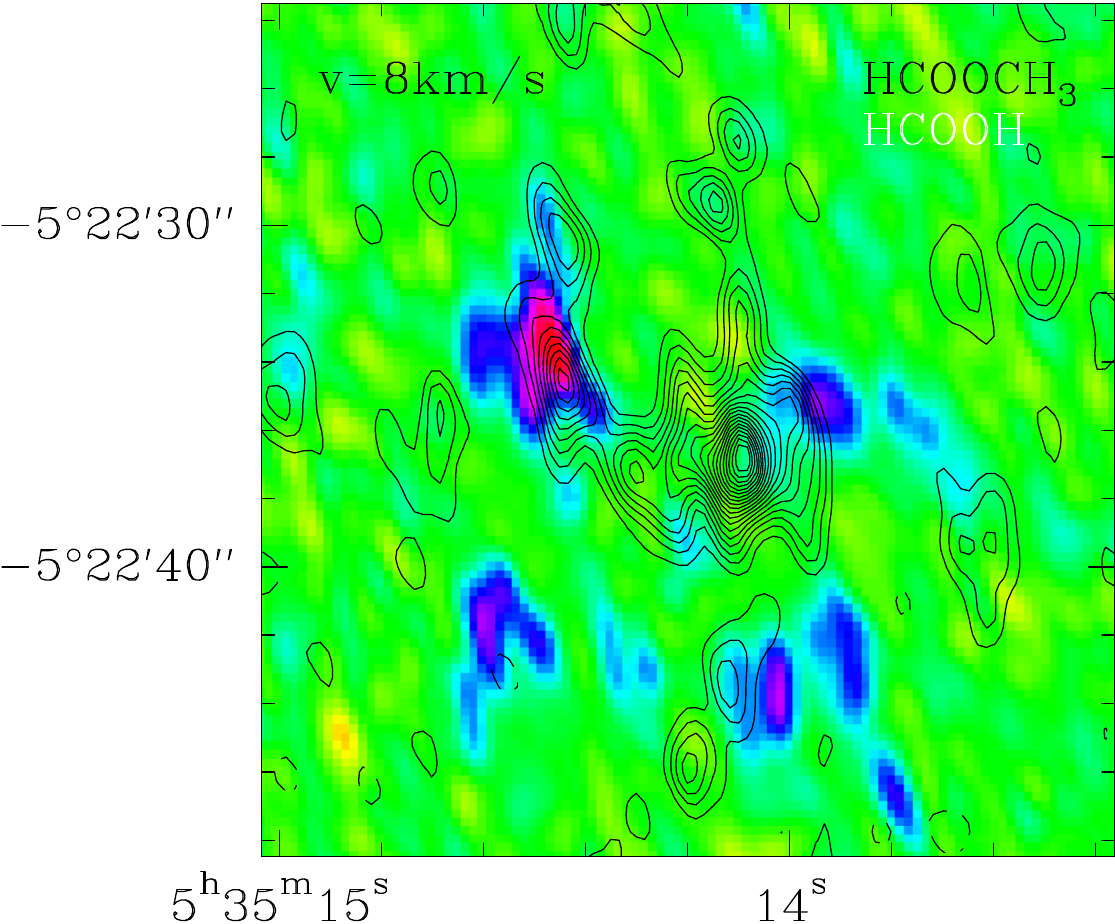}
 Ê\caption{Methyl formate channel map (black contours) at 8~km~s$^{-1}$ (sum of the transitions at 223.465 and 223.500~GHz, E$_u$=50~K) overlaid on the formic acid channel map (colors) at the same velocity for the 223.915~GHz transition (E$_u$=72~K). All data are taken from the same data set (1.79$\arcsec$ $\times$ 0.79$\arcsec$ resolution).}
 ÊÊÊÊÊÊÊÊÊÊÊÊ\label{fig-mf-hcooh-2}%
 ÊÊ\end{figure}

If the molecules would be mainly produced in the gas phase, methyl formate could be produced from formic acid as suggested by \citet{Neill:2011}: the anti-correlation observed between HCOOH and HCOOCH$_{3}$ across most of the Orion-KL region (see Fig. \ref{fig-mf-hcooh-2}) is consistent with their model of recent gas-phase conversion. Two ion--molecule reactions involving the reaction of methanol and formic acid, where one of the reactants is protonated, could be viable interstellar reaction routes to form methyl formate.  Either the methanol, CH$_{3}$OH, or the protonated methanol, CH$_{3}$OH$_{2}$$^{+}$, would then be the common precursor to methyl formate and dimethyl ether.  Indeed, the reaction between protonated methanol and neutral methanol produces protonated dimethyl ether and is considered as an important contributor to interstellar dimethyl ether formation. If methyl formate is efficiently produced in a region by the reaction between HCOOH and CH$_{3}$OH$_{2}$$^{+}$, dimethyl ether is produced as well in the same spatial region, since neutral methanol is also present. 
A prediction of these reactions is the presence of the less stable t-HCOOCH$_{3}$, which is not yet detected in Orion, and the anticorrelation with HCOOH, consumed by the reaction.  Another interesting test would be to detect directly CH$_{3}$OH$_{2}$$^{+}$, once its millimetric spectrum is known. 

%
 Ê
%
%
%

Methanol maps  \citep[see maps in][from the same data set]{Peng:2012} are somewhat different. 
Despite the fact methyl formate and dimethyl ether are chemically related to CH$_{3}$OH, their distributions are not expected to correlate as tightly with methanol as with each other since: i) in the gas phase production scheme,  the methyl formate/dimethyl ether precursor CH$_{3}$OH$_{2}$$^{+}$ can be produced with an efficiency varying across Orion-KL.  In addition CH$_{3}$OH, as the parent species, is partly consumed to some extent (5-20\%) to form methyl formate and dimethyl ether (and could have other reactions); 
ii)  in the ice mantle production scheme, some layering of the mantle is expected \citep[see e.g.][]{Herbst:2009}, and only a fraction of solid methanol closest to the surface can be processed by UV. \\

We suggest, however, another possibility which could be investigated for the gas-phase chemistry. Because the main problem to form HCOOCH$_{3}$ from CH$_{2}$OH$_{2}$$^{+}$ and H$_{2}$CO is an energy barrier, could this obstacle be removed if one considers that the ions are strongly accelerated with respect to neutrals at some places of an MHD shock ? A few km/s would represent enough kinetic energy to overcome the reaction barrier, which  is 128~kJ/mol (15 000~K or 1.2~eV) according to \citet{Horn:2004}.
New quantum calculations and/or laboratory work are required to test this hypothesis.\\

\paragraph{An alternative hypothesis could be: convergence of methyl formate and dimethyl ether abundances with time.\\}

We cannot exclude the convergence with time of the chemical network to a same methyl formate/dimethyl ether ratio value for a  large range of initial conditions and parameters. (Note that such a convergence is not always present; for instance chemical systems may oscillate.) No common precursor is needed in this case. The convergence should be achieved in a time short enough (at least $<$~1-10~Myr , the probable maximum age of molecular clouds in OMC 1). This hypothesis is  less likely in our view (cf the probable consequences of recent shocks).

%
%
\section{Conclusions}

We have studied the distribution of the complex O-bearing molecule dimethyl ether CH$_{3}$OCH$_{3}$ at medium and high angular resolution (6.5$\arcsec$ -- 1$\arcsec$) using various sets of interferometric data from the IRAM Plateau de Bure Interferometer. 
Our main results and conclusions are the following: 
   \begin{enumerate}
      \item
This data set includes 4 well detected lines from E$_u$=19~K to E$_u$=330~K and other partially blended or blended lines.
      \item 
      The most intense emission arises from the Compact Ridge at the methyl formate peak position \citep[MF1,][]{Favre:2011a} ; the second most intense at the Hot Core SW position \citep{Friedel:2008} or MF2 \citep{Favre:2011a}. 
       \item 
       We have used the temperature derived from our previous methyl formate study for the five main peaks and we have deduced CH$_{3}$OCH$_{3}$ column densities assuming LTE. Temperatures cover the range 80 to 170 K, and column densities 1.5$\times$10$^{16}$ to 3.5$\times$10$^{17}$ cm$^{-2}$. 
       \item 
       The abundance of CH$_{3}$OCH$_{3}$ relative to methanol, CH$_{3}$OH, is in the range 0.05 to 0.2 depending on the positions and assuming identical spatial distribution and temperature for both molecules. When observed on the same spatial scale of 3.6$\arcsec$ $\times$  2.3$\arcsec$, the abundance of CH$_{3}$OCH$_{3}$ relative to methyl formate, HCOOCH$_{3}$, is in the range 3.5 to 9.
       \item 
       We observe a very good correlation of the spatial distribution of the methyl formate and dimethyl ether emission. We show that it is most likely due to a correlation of their abundances. 

       \item 
     The dimethyl ether emission follows the 2.12~$\mu m$ H$_{2}$ distribution, as does the methyl formate distribution. Shocks seem to be related to the presence of these species, possibly because of the release of molecules from grain mantles.
         
          \item
           A common precursor to dimethyl ether and methyl formate appears the simplest explanation to the observed correlation. In the gas phase,  the precursor CH$_{3}$OH$_{2}$$^{+}$ would react respectively with CH$_{3}$OH and, as suggested by \citet{Neill:2011}, with HCOOH. The reaction of protonated methanol CH$_{3}$OH$_{2}$$^{+}$ with H$_{2}$CO and CH$_{3}$OH is excluded by quantum calculations in normal interstellar conditions. We speculate whether the reaction of high speed CH$_{3}$OH$_{2}$$^{+}$ ions (a few km/s to $>$ 20~km/s) might change this conclusion. 
           
           On the other hand, if methyl formate and dimethyl ether are already produced in the ice mantle, CH$_{3}$OH and/or methoxy radical CH$_{3}$O seem to be the common precursor.

       \item 
   We observe a different distribution of ethanol C$_{2}$H$_{5}$OH (and ethylene glycol CH$_{2}$OHCH$_{2}$OH, in prep.). In the grain mantle chemistry scheme this could result from different production rates or schemes of CH$_{2}$OH vs CH$_{3}$O radicals across Orion-KL.   
    
    Furthermore we confirm the anti-correlation of methyl formate with formic acid, HCOOH, found by \citet{Neill:2011}. This in favor of their model of  production of methyl formate from gas phase HCOOH and CH$_{3}$OH$_{2}$$^{+}$  
where CH$_{3}$OH$_{2}$$^{+}$ is formed from methanol released from grains.

  \end{enumerate}

High resolution mapping brings new insight to the comparison of complex organic molecules and the understanding of  their formation. The HCOOCH$_{3}$ and CH$_{3}$OCH$_{3}$ tight correlation, and the different behavior of ethanol we observe are a strong constraint for future chemical models.

We identify  the need for the  following key studies to constrain further the chemistry of dimethyl ether and O-bearing complex species: search for mantle species produced by  the CH$_{2}$OH radical; search for CH$_{3}$OH$_{2}$$^{+}$; maps of t-HCOOCH$_{3}$ and HCOOH; theoretical/laboratory studies of protonated methanol reaction at high speed (shocked) on H$_{2}$CO and CH$_{3}$OH. 
The conditions leading to the convergence of the dimethyl ether/methyl formate ratio in a complex chemical network could also be investigated.

Shock related molecule release from grains and  shock induced chemistry of complex molecules need to be modeled. The issue of the efficiency or lack of such chemical conditions to transform molecules in a short time lapse ($< $10$^3$~yr) is critical in the discussion of the merits of the two methyl formate and dimethyl ether production schemes (gas phase production vs simple release); this will be addressed in a future work.

%
%
\begin{acknowledgements}
      This research has made use of the SIMBAD database, operated at the CDS, Strasbourg, France, and of the Splatalogue database \citep[http://www.splatalogue.net,][]{Remijan:2007}.
      This work was supported by the CNRS national programs PCMI (Physics and Chemistry of the Interstellar Medium) and GDR Exobiology.
      We thank J. Cernicharo for his IRAM 30m spectra. We thank the referee and the editor for raising interesting issues.

\end{acknowledgements}

%
%
\bibliographystyle{aa}
\bibliography{biblio}

\begin{thebibliography}{63}
\expandafter\ifx\csname natexlab\endcsname\relax\def\natexlab#1{#1}\fi

\bibitem[{{Allen} \& {Burton}(1993)}]{Allen:1993}
{Allen}, D.~A. \& {Burton}, M.~G. 1993, \nat, 363, 54

\bibitem[{{Belloche} {et~al.}(2009){Belloche}, {Garrod}, {M{\"u}ller},
  {Menten}, {Comito}, \& {Schilke}}]{Belloche:2009}
{Belloche}, A., {Garrod}, R.~T., {M{\"u}ller}, H.~S.~P., {et~al.} 2009, \aap,
  499, 215

\bibitem[{{Bennett} \& {Kaiser}(2007)}]{Bennett:2007}
{Bennett}, C.~J. \& {Kaiser}, R.~I. 2007, \apj, 661, 899

\bibitem[{{Beuther} {et~al.}(2005){Beuther}, {Zhang}, {Greenhill}, {Reid},
  {Wilner}, {Keto}, {Shinnaga}, {Ho}, {Moran}, {Liu}, \&
  {Chang}}]{Beuther:2005}
{Beuther}, H., {Zhang}, Q., {Greenhill}, L.~J., {et~al.} 2005, \apj, 632, 355

\bibitem[{{Beuther} {et~al.}(2006){Beuther}, {Zhang}, {Reid}, {Hunter},
  {Gurwell}, {Wilner}, {Zhao}, {Shinnaga}, {Keto}, {Ho}, {Moran}, \&
  {Liu}}]{Beuther:2006}
{Beuther}, H., {Zhang}, Q., {Reid}, M.~J., {et~al.} 2006, \apj, 636, 323

\bibitem[{{Bisschop} {et~al.}(2007){Bisschop}, {J{\o}rgensen}, {van Dishoeck},
  \& {de Wachter}}]{Bisschop:2007}
{Bisschop}, S.~E., {J{\o}rgensen}, J.~K., {van Dishoeck}, E.~F., \& {de
  Wachter}, E.~B.~M. 2007, \aap, 465, 913

\bibitem[{{Blake}(1988)}]{Blake:1988}
{Blake}, G.~A. 1988, in Lecture Notes in Physics, Berlin Springer Verlag, Vol.
  315, Molecular Clouds, Milky-Way and External Galaxies, ed. {R.~L.~Dickman,
  R.~L.~Snell, \& J.~S.~Young}, 132

\bibitem[{{Blake} {et~al.}(1996){Blake}, {Mundy}, {Carlstrom}, {Padin},
  {Scott}, {Scoville}, \& {Woody}}]{Blake:1996}
{Blake}, G.~A., {Mundy}, L.~G., {Carlstrom}, J.~E., {et~al.} 1996, \apjl, 472,
  L49+

\bibitem[{{Blake} {et~al.}(1987){Blake}, {Sutton}, {Masson}, \&
  {Phillips}}]{Blake:1987}
{Blake}, G.~A., {Sutton}, E.~C., {Masson}, C.~R., \& {Phillips}, T.~G. 1987,
  \apj, 315, 621

\bibitem[{{Caselli} {et~al.}(1993){Caselli}, {Hasegawa}, \&
  {Herbst}}]{Caselli:1993}
{Caselli}, P., {Hasegawa}, T.~I., \& {Herbst}, E. 1993, \apj, 408, 548

\bibitem[{{Cernicharo} {et~al.}(2012){Cernicharo}, {Marcelino}, {Roueff},
  {Gerin}, {Jim{\'e}nez-Escobar}, \& {Mu{\~n}oz Caro}}]{Cernicharo:2012}
{Cernicharo}, J., {Marcelino}, N., {Roueff}, E., {et~al.} 2012, \apjl, 759, L43

\bibitem[{{Charnley} {et~al.}(1995){Charnley}, {Kress}, {Tielens}, \&
  {Millar}}]{Charnley:1995}
{Charnley}, S.~B., {Kress}, M.~E., {Tielens}, A.~G.~G.~M., \& {Millar}, T.~J.
  1995, \apj, 448, 232

\bibitem[{{Clark}(1980)}]{Clark:1980}
{Clark}, B.~G. 1980, \aap, 89, 377

\bibitem[{{Clark} {et~al.}(1979){Clark}, {Lovas}, \& {Johnson}}]{Clark:1979}
{Clark}, F.~O., {Lovas}, F.~J., \& {Johnson}, D.~R. 1979, \apj, 229, 553

\bibitem[{{Comito} {et~al.}(2005){Comito}, {Schilke}, {Phillips}, {Lis},
  {Motte}, \& {Mehringer}}]{Comito:2005}
{Comito}, C., {Schilke}, P., {Phillips}, T.~G., {et~al.} 2005, \apjs, 156, 127

\bibitem[{{Eisner} {et~al.}(2008){Eisner}, {Plambeck}, {Carpenter}, {Corder},
  {Qi}, \& {Wilner}}]{Eisner:2008}
{Eisner}, J.~A., {Plambeck}, R.~L., {Carpenter}, J.~M., {et~al.} 2008, \apj,
  683, 304

\bibitem[{{Endres} {et~al.}(2009){Endres}, {Drouin}, {Pearson}, {M{\"u}ller},
  {Lewen}, {Schlemmer}, \& {Giesen}}]{Endres:2009}
{Endres}, C.~P., {Drouin}, B.~J., {Pearson}, J.~C., {et~al.} 2009, \aap, 504,
  635

\bibitem[{{Favre} {et~al.}(2011{\natexlab{a}}){Favre}, {Despois}, {Brouillet},
  {Baudry}, {Combes}, {Gu\'elin}, {Wootten}, \& {Wlodarczak}}]{Favre:2011a}
{Favre}, C., {Despois}, D., {Brouillet}, N., {et~al.} 2011{\natexlab{a}}, \aap,
  532, 32

\bibitem[{{Favre} {et~al.}(2011{\natexlab{b}}){Favre}, {Wootten}, {Remijan},
  {Brouillet}, {Wilson}, {Despois}, \& {Baudry}}]{Favre:2011b}
{Favre}, C., {Wootten}, H.~A., {Remijan}, A.~J., {et~al.} 2011{\natexlab{b}},
  \apjl, 739, L12

\bibitem[{{Friedel} \& {Snyder}(2008)}]{Friedel:2008}
{Friedel}, D.~N. \& {Snyder}, L.~E. 2008, \apj, 672, 962

\bibitem[{{Garrod} {et~al.}(2008){Garrod}, {Weaver}, \& {Herbst}}]{Garrod:2008}
{Garrod}, R.~T., {Weaver}, S.~L.~W., \& {Herbst}, E. 2008, \apj, 682, 283

\bibitem[{{Goddi} {et~al.}(2009){Goddi}, {Greenhill}, {Humphreys}, {Matthews},
  {Tan}, \& {Chandler}}]{Goddi:2009a}
{Goddi}, C., {Greenhill}, L.~J., {Humphreys}, E.~M.~L., {et~al.} 2009, \apj,
  691, 1254

\bibitem[{{Goddi} {et~al.}(2011){Goddi}, {Humphreys}, {Greenhill}, {Chandler},
  \& {Matthews}}]{Goddi:2011}
{Goddi}, C., {Humphreys}, E.~M.~L., {Greenhill}, L.~J., {Chandler}, C.~J., \&
  {Matthews}, L.~D. 2011, \apj, 728, 15

\bibitem[{{G{\'o}mez} {et~al.}(2005){G{\'o}mez}, {Rodr{\'{\i}}guez}, {Loinard},
  {Lizano}, {Poveda}, \& {Allen}}]{Gomez:2005}
{G{\'o}mez}, L., {Rodr{\'{\i}}guez}, L.~F., {Loinard}, L., {et~al.} 2005, \apj,
  635, 1166

\bibitem[{{Gredel} {et~al.}(1989){Gredel}, {Lepp}, {Dalgarno}, \&
  {Herbst}}]{Gredel:1989}
{Gredel}, R., {Lepp}, S., {Dalgarno}, A., \& {Herbst}, E. 1989, \apj, 347, 289

\bibitem[{{Gu{\'e}lin} {et~al.}(2008){Gu{\'e}lin}, {Brouillet}, {Cernicharo},
  {Combes}, \& {Wooten}}]{Guelin:2008}
{Gu{\'e}lin}, M., {Brouillet}, N., {Cernicharo}, J., {Combes}, F., \& {Wooten},
  A. 2008, \apss, 313, 45

\bibitem[{{Herbst} \& {van Dishoeck}(2009)}]{Herbst:2009}
{Herbst}, E. \& {van Dishoeck}, E.~F. 2009, \araa, 47, 427

\bibitem[{{Hirota} {et~al.}(2007){Hirota}, {Bushimata}, {Choi}, {Honma},
  {Imai}, {Iwadate}, {Jike}, {Kameno}, {Kameya}, {Kamohara}, {Kan-Ya},
  {Kawaguchi}, {Kijima}, {Kim}, {Kobayashi}, {Kuji}, {Kurayama}, {Manabe},
  {Maruyama}, {Matsui}, {Matsumoto}, {Miyaji}, {Nagayama}, {Nakagawa},
  {Nakamura}, {Oh}, {Omodaka}, {Oyama}, {Sakai}, {Sasao}, {Sato}, {Sato},
  {Shibata}, {Shintani}, {Tamura}, {Tsushima}, \& {Yamashita}}]{Hirota:2007}
{Hirota}, T., {Bushimata}, T., {Choi}, Y.~K., {et~al.} 2007, \pasj, 59, 897

\bibitem[{{Horn} {et~al.}(2004){Horn}, {M{\o}llendal}, {Sekiguchi}, {Uggerud},
  {Roberts}, {Herbst}, {Viggiano}, \& {Fridgen}}]{Horn:2004}
{Horn}, A., {M{\o}llendal}, H., {Sekiguchi}, O., {et~al.} 2004, \apj, 611, 605

\bibitem[{{Johansson} {et~al.}(1985){Johansson}, {Andersson}, {Elder},
  {Friberg}, {Hjalmarson}, {Hoglund}, {Olofsson}, {Rydbeck}, \&
  {Irvine}}]{Johansson:1985}
{Johansson}, L.~E.~B., {Andersson}, C., {Elder}, J., {et~al.} 1985, \aaps, 60,
  135

\bibitem[{{Johansson} {et~al.}(1984){Johansson}, {Andersson}, {Ellder},
  {Friberg}, {Hjalmarson}, {Hoglund}, {Irvine}, {Olofsson}, \&
  {Rydbeck}}]{Johansson:1984}
{Johansson}, L.~E.~B., {Andersson}, C., {Ellder}, J., {et~al.} 1984, \aap, 130,
  227

\bibitem[{{J{\o}rgensen} {et~al.}(2011){J{\o}rgensen}, {Bourke}, {Nguyen
  Luong}, \& {Takakuwa}}]{Jorgensen:2011}
{J{\o}rgensen}, J.~K., {Bourke}, T.~L., {Nguyen Luong}, Q., \& {Takakuwa}, S.
  2011, \aap, 534

\bibitem[{{Kim} {et~al.}(2008){Kim}, {Hirota}, {Honma}, {Kobayashi},
  {Bushimata}, {Choi}, {Imai}, {Iwadate}, {Jike}, {Kameno}, {Kameya},
  {Kamohara}, {Kan-Ya}, {Kawaguchi}, {Kuji}, {Kurayama}, {Manabe}, {Matsui},
  {Matsumoto}, {Miyaji}, {Nagayama}, {Nakagawa}, {Oh}, {Omodaka}, {Oyama},
  {Sakai}, {Sasao}, {Sato}, {Sato}, {Shibata}, {Tamura}, \&
  {Yamashita}}]{Kim:2008}
{Kim}, M.~K., {Hirota}, T., {Honma}, M., {et~al.} 2008, \pasj, 60, 991

\bibitem[{{Laas} {et~al.}(2011){Laas}, {Garrod}, {Herbst}, \& {Widicus
  Weaver}}]{Laas:2011}
{Laas}, J.~C., {Garrod}, R.~T., {Herbst}, E., \& {Widicus Weaver}, S.~L. 2011,
  \apj, 728, 71

\bibitem[{{Lee} \& {Cho}(2002)}]{Lee:2002}
{Lee}, C.~W. \& {Cho}, S.-H. 2002, {Journal of Korean Astronomical Society},
  35, 187

\bibitem[{{Lovas} {et~al.}(1979){Lovas}, {Johnson}, \& {Snyder}}]{Lovas:1979}
{Lovas}, F.~J., {Johnson}, D.~R., \& {Snyder}, L.~E. 1979, \apjs, 41, 451

\bibitem[{{Menten} {et~al.}(2007){Menten}, {Reid}, {Forbrich}, \&
  {Brunthaler}}]{Menten:2007}
{Menten}, K.~M., {Reid}, M.~J., {Forbrich}, J., \& {Brunthaler}, A. 2007, \aap,
  474, 515

\bibitem[{{Minh} {et~al.}(1993){Minh}, {Ohishi}, {Roh}, {Ishiguro}, \&
  {Irvine}}]{Minh:1993}
{Minh}, Y.~C., {Ohishi}, M., {Roh}, D.~G., {Ishiguro}, M., \& {Irvine}, W.~M.
  1993, \apj, 411, 773

\bibitem[{{M{\"u}ller} {et~al.}(2005){M{\"u}ller}, {Schl{\"o}der}, {Stutzki},
  \& {Winnewisser}}]{Muller:2005}
{M{\"u}ller}, H.~S.~P., {Schl{\"o}der}, F., {Stutzki}, J., \& {Winnewisser}, G.
  2005, Journal of Molecular Structure, 742, 215

\bibitem[{{M{\"u}ller} {et~al.}(2001){M{\"u}ller}, {Thorwirth}, {Roth}, \&
  {Winnewisser}}]{Muller:2001}
{M{\"u}ller}, H.~S.~P., {Thorwirth}, S., {Roth}, D.~A., \& {Winnewisser}, G.
  2001, \aap, 370, L49

\bibitem[{{Neill} {et~al.}(2011){Neill}, {Steber}, {Muckle}, {Zaleski},
  {Lattanzi}, {Spezzano}, {McCarthy}, {Remijan}, {Friedel}, {Widicus Weaver},
  \& {Pate}}]{Neill:2011}
{Neill}, J.~L., {Steber}, A.~L., {Muckle}, M.~T., {et~al.} 2011, {Journal of
  Physical Chemistry A}, 115, 6472

\bibitem[{{Neufeld} \& {Dalgarno}(1989)}]{Neufeld:1989}
{Neufeld}, D.~A. \& {Dalgarno}, A. 1989, \apj, 340, 869

\bibitem[{{Nissen} {et~al.}(2012){Nissen}, {Cunningham}, {Gustafsson}, {Bally},
  {Lemaire}, {Favre}, \& {Field}}]{Nissen:2012}
{Nissen}, H.~D., {Cunningham}, N.~J., {Gustafsson}, M., {et~al.} 2012, \aap,
  540, A119

\bibitem[{{{\"O}berg} {et~al.}(2010){{\"O}berg}, {Bottinelli}, {Jorgensen}, \&
  {van Dishoeck}}]{Oberg:2010}
{{\"O}berg}, K.~I., {Bottinelli}, S., {Jorgensen}, J.~K., \& {van Dishoeck},
  E.~F. 2010, \apj, 716, 825

\bibitem[{{{\"O}berg} {et~al.}(2009){{\"O}berg}, {Garrod}, {van Dishoeck}, \&
  {Linnartz}}]{Oberg:2009}
{{\"O}berg}, K.~I., {Garrod}, R.~T., {van Dishoeck}, E.~F., \& {Linnartz}, H.
  2009, \aap, 504, 891

\bibitem[{{Peeters} {et~al.}(2006){Peeters}, {Rodgers}, {Charnley},
  {Schriver-Mazzuoli}, {Schriver}, {Keane}, \& {Ehrenfreund}}]{Peeters:2006}
{Peeters}, Z., {Rodgers}, S.~D., {Charnley}, S.~B., {et~al.} 2006, \aap, 445,
  197

\bibitem[{{Peng} {et~al.}(2012){Peng}, {Despois}, {Brouillet}, {Parise}, \&
  {Baudry}}]{Peng:2012}
{Peng}, T.~C., {Despois}, D., {Brouillet}, N., {Parise}, B., \& {Baudry}, A.
  2012, \aap, in press

\bibitem[{{Persson} {et~al.}(2007){Persson}, {Olofsson}, {Koning}, {Bergman},
  {Bernath}, {Black}, {Frisk}, {Geppert}, {Hasegawa}, {Hjalmarson}, {Kwok},
  {Larsson}, {Lecacheux}, {Nummelin}, {Olberg}, {Sandqvist}, \&
  {Wirstr{\"o}m}}]{Persson:2007}
{Persson}, C.~M., {Olofsson}, A.~O.~H., {Koning}, N., {et~al.} 2007, \aap, 476,
  807

\bibitem[{{Prasad} \& {Tarafdar}(1983)}]{Prasad:1983}
{Prasad}, S.~S. \& {Tarafdar}, S.~P. 1983, \apj, 267, 603

\bibitem[{{Remijan} {et~al.}(2008){Remijan}, {Leigh}, {Markwick-Kemper}, \&
  {Turner}}]{Remijan:2008}
{Remijan}, A.~J., {Leigh}, D.~P., {Markwick-Kemper}, A.~J., \& {Turner}, B.~E.
  2008, ArXiv e-prints

\bibitem[{{Remijan} {et~al.}(2007){Remijan}, {Markwick-Kemper}, \& {ALMA
  Working Group on Spectral Line Frequencies}}]{Remijan:2007}
{Remijan}, A.~J., {Markwick-Kemper}, A., \& {ALMA Working Group on Spectral
  Line Frequencies}. 2007, in Bulletin of the American Astronomical Society,
  Vol.~39, American Astronomical Society Meeting Abstracts, 132.11

\bibitem[{{Requena-Torres} {et~al.}(2008){Requena-Torres},
  {Mart{\'{\i}}n-Pintado}, {Mart{\'{\i}}n}, \& {Morris}}]{Requena-Torres:2008}
{Requena-Torres}, M.~A., {Mart{\'{\i}}n-Pintado}, J., {Mart{\'{\i}}n}, S., \&
  {Morris}, M.~R. 2008, \apj, 672, 352

\bibitem[{{Requena-Torres} {et~al.}(2006){Requena-Torres},
  {Mart{\'{\i}}n-Pintado}, {Rodr{\'{\i}}guez-Franco}, {Mart{\'{\i}}n},
  {Rodr{\'{\i}}guez-Fern{\'a}ndez}, \& {de Vicente}}]{Requena-Torres:2006}
{Requena-Torres}, M.~A., {Mart{\'{\i}}n-Pintado}, J.,
  {Rodr{\'{\i}}guez-Franco}, A., {et~al.} 2006, \aap, 455, 971

\bibitem[{{Rodr{\'{\i}}guez} {et~al.}(2005){Rodr{\'{\i}}guez}, {Poveda},
  {Lizano}, \& {Allen}}]{Rodriguez:2005}
{Rodr{\'{\i}}guez}, L.~F., {Poveda}, A., {Lizano}, S., \& {Allen}, C. 2005,
  \apjl, 627, L65

\bibitem[{{Sandstrom} {et~al.}(2007){Sandstrom}, {Peek}, {Bower}, {Bolatto}, \&
  {Plambeck}}]{Sandstrom:2007}
{Sandstrom}, K.~M., {Peek}, J.~E.~G., {Bower}, G.~C., {Bolatto}, A.~D., \&
  {Plambeck}, R.~L. 2007, \apj, 667, 1161

\bibitem[{{Schilke} {et~al.}(2001){Schilke}, {Benford}, {Hunter}, {Lis}, \&
  {Phillips}}]{Schilke:2001}
{Schilke}, P., {Benford}, D.~J., {Hunter}, T.~R., {Lis}, D.~C., \& {Phillips},
  T.~G. 2001, \apjs, 132, 281

\bibitem[{{Schilke} {et~al.}(1997){Schilke}, {Groesbeck}, {Blake}, \&
  {Phillips}}]{Schilke:1997}
{Schilke}, P., {Groesbeck}, T.~D., {Blake}, G.~A., \& {Phillips}, T.~G. 1997,
  \apjs, 108, 301

\bibitem[{{Snyder} {et~al.}(1974){Snyder}, {Buhl}, {Schwartz}, {Clark},
  {Johnson}, {Lovas}, \& {Giguere}}]{Snyder:1974}
{Snyder}, L.~E., {Buhl}, D., {Schwartz}, P.~R., {et~al.} 1974, \apjl, 191, L79

\bibitem[{{Sutton} {et~al.}(1995){Sutton}, {Peng}, {Danchi}, {Jaminet},
  {Sandell}, \& {Russell}}]{Sutton:1995}
{Sutton}, E.~C., {Peng}, R., {Danchi}, W.~C., {et~al.} 1995, \apjs, 97, 455

\bibitem[{{Turner}(1989)}]{Turner:1989}
{Turner}, B.~E. 1989, \apjs, 70, 539

\bibitem[{{Turner}(1991)}]{Turner:1991}
{Turner}, B.~E. 1991, \apjs, 76, 617

\bibitem[{{Zapata} {et~al.}(2009){Zapata}, {Schmid-Burgk}, {Ho},
  {Rodr{\'{\i}}guez}, \& {Menten}}]{Zapata:2009}
{Zapata}, L.~A., {Schmid-Burgk}, J., {Ho}, P.~T.~P., {Rodr{\'{\i}}guez}, L.~F.,
  \& {Menten}, K.~M. 2009, \apjl, 704, L45

\bibitem[{{Ziurys} \& {McGonagle}(1993)}]{Ziurys:1993}
{Ziurys}, L.~M. \& {McGonagle}, D. 1993, \apjs, 89, 155

\end{thebibliography}

\end{document}